\documentclass[aps,prl,reprint,superscriptaddress,nofootinbib,floatfix]{revtex4-2}

\usepackage{amsmath,amssymb,bm}
\usepackage{graphicx}
\usepackage{microtype}
\usepackage[hidelinks]{hyperref}
\usepackage{physics}

\begin{document}

\title{Criticality Concentrates Extensive Sensing Information into a Local Readout}


\author{Peng Wang}
\email{wp002005@163.com}
\altaffiliation{These authors contributed equally to this work.}
\affiliation{School of Computer and Artificial Intelligence, Southwest Minzu University, Chengdu 610225, China}
\affiliation{Centre for Quantum Technologies, National University of Singapore, Singapore 117543, Singapore}

\author{Yu-Xuan Zhang}
\email{2411915@mail.nankai.edu.cn}
\altaffiliation{These authors contributed equally to this work.}
\affiliation{School of Physics, Nankai University, Tianjin 300071, China}
\affiliation{Centre for Quantum Technologies, National University of Singapore, Singapore 117543, Singapore}

\author{Hai-Tao Ding}
\email{htding.9@nus.edu.sg}
\affiliation{Centre for Quantum Technologies, National University of Singapore, Singapore 117543, Singapore}
\affiliation{MajuLab, CNRS-UNS-NUS-NTU International Joint Research Unit, Singapore UMI 3654, Singapore}

\author{Leong-Chuan Kwek}
\email{kwekleongchuan@nus.edu.sg}
\affiliation{Centre for Quantum Technologies, National University of Singapore, Singapore 117543, Singapore}
\affiliation{MajuLab, CNRS-UNS-NUS-NTU International Joint Research Unit, Singapore UMI 3654, Singapore}
\affiliation{National Institute of Education, Nanyang Technological University, Singapore 637616, Singapore}

\begin{abstract}
We show that criticality concentrates extensive sensing information into a local trajectory. In the rapid-readout limit of an all-to-all dissipative feedback network, a single binary trajectory has a Fisher-information rate linear in $N$ for a common weak field, as does the full record. Critical slowing does not remove this gain because the local
weight of the slow mode decreases with $N$. An analytic bound guarantees that the local record retains more than half of the full information across the entire channel-parameter range. We also identify how parameter drives and observation dynamics determine local access to collective information beyond permutation symmetry. Furthermore, our sparse-readout theorem establishes a
$k^{-1}$ decay of fractional information loss: only
$\Theta(\varepsilon^{-1})$ trajectories are needed to retain
a fraction $1-\varepsilon$ of the full information rate. Any growing readout set, even of vanishing
density, asymptotically recovers the full common-field
Fisher-information rate.
\end{abstract}

\maketitle

\paragraph{Introduction.---}
Interacting sensors can encode information in collective degrees of freedom that are difficult to access locally. The precision of an estimator depends both on the probe and on the outputs it retains~\cite{Degen2017,Braun2018,Endres2009Receptor,Smiga2023Empirical,Yang2026Noisy,Blinova2026Output}. Criticality sharpens this distinction: enhanced susceptibility improves parameter sensitivity~\cite{Zanardi2008,Frerot2018,Mihailescu2026Tutorial,Montenegro2025Review}, while slow relaxation can limit the information acquired per unit time~\cite{Hohenberg1977,Rams2018,Garbe2020,Gietka2021Adiabatic,Garbe2022FullyConnected,Alushi2024Optimality,Gyhm2026Scaling,Lee2025Timescales}. A local sensing advantage must therefore survive the temporal correlations of the measured record. Finite-time and prior-uncertainty constraints also motivate nonadiabatic~\cite{Shimshi2026Beyond} and adaptive-feedback~\cite{Salvia2023Feedback} protocols, respectively.

Collective and dissipative critical sensing~\cite{Chu2021,Macieszczak2016,Raghunandan2018,Ilias2022,Fernandez2017,DiCandia2023Parametric,Alushi2025Collective,Ilias2024Ion}, including experimental implementations~\cite{Ding2022,Beaulieu2025Resonator,Lu2026Robust,Wang2026Flipping}, local thermometry and reduced-state metrology~\cite{DePasquale2016,Bina2016,Adani2024,Mihailescu2023Thermometry,Mihailescu2024Impurity}, and sensing under partial accessibility~\cite{Mishra2021Partial,Montenegro2022Optomech,Zhao2026Partial} establish the value of critical response and restricted observations. Continuous or sequential measurement~\cite{Gammelmark2014,Kiilerich2014,Radaelli2023,Yang2023PRX,Yang2023Sequential,Gammelmark2013Bayesian,Kiilerich2016,OConnor2024Rates,Radaelli2026Jump,Montenegro2022Sequential,Mattes2025Classical,Cabot2024TimeCrystal,OConnor2026Monitored} and Gaussian output-field analyses~\cite{Genoni2017Gaussian,Yokomizo2026Gaussian} further address parameter estimation from time-dependent outputs. Cooperative receptor models likewise connect response enhancement to noise and response speed~\cite{Skoge2011Cooperativity,Skoge2013Nonequilibrium,Hathcock2024Dissipation}. Multicellular feedback models also relate a local readout to spatially averaged signals, but find a slowing-down penalty in stochastic-signal information rates~\cite{Vennettilli2020}. Despite these advances, how much of the full-network Fisher-information rate can be extracted from a fixed or sparse set of local temporal records remains unclear. For a time-independent common field, we address two related trajectory-level resource questions: Can a single continuously observed binary node retain a finite fraction of the full-network Fisher-information rate? More generally, how many local trajectories are needed to recover nearly all of that rate? Because the unobserved nodes render a local record non-Markovian, neither static susceptibility nor full-network Fisher information alone resolves these questions.

\begin{figure*}[t]
\centering
\includegraphics[width=0.9\textwidth]{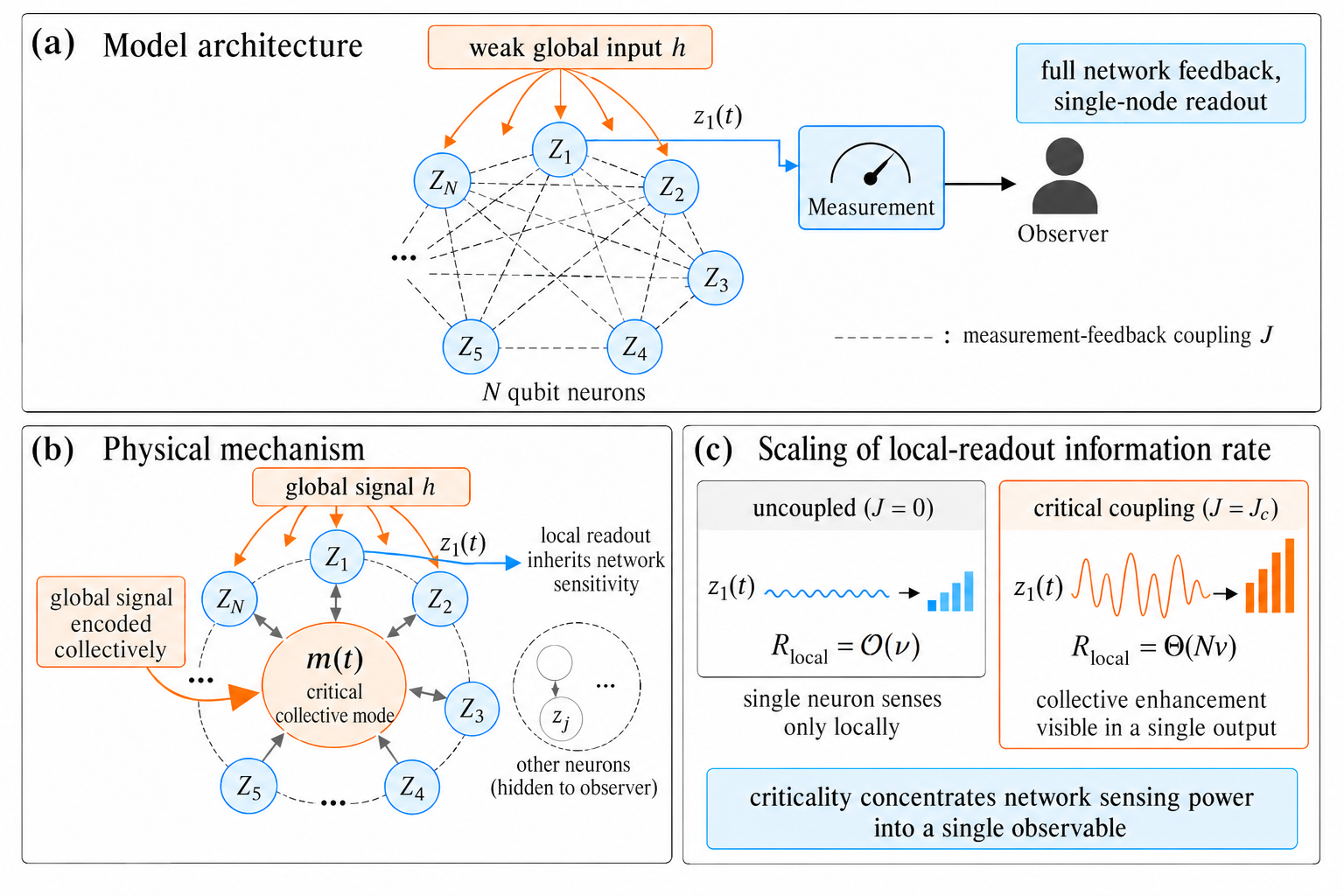}
\caption{Critical local readout in an all-to-all network. (a) A common field $h$ drives $N$ dissipative qubit neurons coupled by classical measurement feedback. The controller uses the network record; the estimator retains only $z_1(t)$. (b) The tagged output inherits collective sensitivity. (c) Its information rate changes from $\mathcal{O}(\nu)$ without coupling to $\Theta(N\nu)$ at criticality, retaining the finite fraction in Eq.~\eqref{eq:central}.}
\label{fig:architecture}
\end{figure*}

Here we prove such retention in a dissipative all-to-all feedback network driven by a common weak field. The estimator sees one tagged trajectory, while the controller uses network-wide measurements. Ref.~\cite{Wang2026Bound} establishes an $\mathcal{O}(\nu)$ Fisher-information-rate benchmark for a single
uncoupled, clamped-input neuron; in contrast, critical coupling here yields
$R_{\mathrm{loc}} = \Theta(N\nu)$ from one observed node of a growing network(Fig.~\ref{fig:architecture}).
We establish the microscopic stationary rate, including its coefficient, by likelihood convergence and a long-time block comparison ( Sec.~S2, ~\cite{sm}). This identifies how an extensive information budget becomes locally accessible. We further prove that the fractional information loss decays as $1/k$
for $k$ retained trajectories, so that any growing readout set,
even one of vanishing density, asymptotically recovers the full
common-field information rate. We also derive visibility criteria for spatially varying drives, general Gaussian networks, and nonlinear observation channels (Sec.~S3, ~\cite{sm}).

The tagged susceptibility grows as $\chi_1\sim N^{1/2}$ and the collective lifetime as $\tau_c\sim N^{1/2}/\nu$. The slow mode's local correlation weight shrinks as $N^{-1/2}$, leaving finite integrated local noise. Consequently,
\begin{equation}
\eta_s\equiv\lim_{N\to\infty}\frac{R_{\rm loc}}{R_{\rm full}}
=2\mathcal J(\rho_s)\ge\eta_s^{\rm mean}>0,
\label{eq:central}
\end{equation}
where $\mathcal J$ is the information rate of the limiting observation channel and $\eta_s^{\rm mean}$ is a time-average lower bound, both specified below. All rates are evaluated at $h=0$ and $J=J_c$, with $s$ and $\nu$ fixed, in the rapid-readout limit. Physical observation time tends to infinity at each fixed $N$ before $N\to\infty$.

\paragraph{Dissipative feedback network.---}
The partial-reset local channel~\cite{Zhang2026,Wang2026Bound} has amplitude-retention parameter $0\le s<1$ and relaxation scale
\begin{equation}
\kappa=\nu(1-s).
\end{equation}
In the rapid-readout measurement-feedback protocol~\cite{Wang2026Ordering}, simultaneous $Z$ readout produces $\bm z=(z_1,\ldots,z_N)$, $z_i=\pm1$. The field and single-spin flip rates are
\begin{equation}
a_i(\bm z;h)=h+\frac{J}{N-1}\sum_{j\ne i}z_j,
\label{eq:field}
\end{equation}
\begin{equation}
w_i(\bm z;h)=\frac{\kappa}{2}
[1-z_i\tanh a_i][1-s z_i\tanh a_i].
\label{eq:rates}
\end{equation}
These rates originate from a dissipative qubit channel, but the observed process and feedback are classical. No entangling interaction is required. Ref.~\cite{Zhang2026}  concerns dephasing-induced stationary ordering and Ref.~\cite{Wang2026Ordering} cadence-controlled ordering; here, instead, the inference problem retains only one node and asks how much of the complete network’s trajectory Fisher information remains locally accessible. The estimator observes
\begin{equation}
Y(t)=z_1(t).
\end{equation}
Permutation symmetry reduces the hidden dynamics exactly to $2N$ states: the tagged sign and the number of positive hidden spins.

\paragraph{Critical finite-size scaling.---}
For magnetization $m=N^{-1}\sum_i z_i$, the large-$N$ drift is
\begin{equation}
\begin{aligned}
\dot m=\kappa\bigl[&(1+s)\tanh(h+Jm)\\
&-\bigl(1+s\tanh^2(h+Jm)\bigr)m\bigr].
\end{aligned}
\end{equation}
Linearization at $m=h=0$ gives the ordering threshold
\begin{equation}
J_c=\frac{1}{1+s}.
\end{equation}
The critical drift and noise expansion~\cite{Binder1981,Hohenberg1977} is
\begin{equation}
\begin{aligned}
 dm={}&\kappa\bigl[(1+s)h-c_s m^3\\
 &\quad+O(m^5,hm^2,h^3,N^{-1})\bigr]dt
 +\sqrt{\frac{2\kappa}{N}}\,dW_t,
\end{aligned}
\label{eq:criticalm}
\end{equation}
with cubic restoring coefficient
\begin{equation}
c_s=\frac{s+1/3}{(1+s)^2}.
\label{eq:cs}
\end{equation}
At zero field the stationary critical density is
\begin{equation}
P_N(m)\propto\exp\left[-\frac{N c_s}{4}m^4+o(1)\right].
\end{equation}
Thus $m=O(N^{-1/4})$ and $\tau_c=O(\sqrt N/\nu)$. Exchangeability and the stationary weights give
\begin{equation}
\begin{aligned}
\chi_1&\equiv\left.\partial_h\langle z_1\rangle\right|_0
=A_s\sqrt N+O(1),\\
A_s&=\frac{2(1+s)}{\sqrt{c_s}}
\frac{\Gamma(3/4)}{\Gamma(1/4)}.
\end{aligned}
\label{eq:chi}
\end{equation}

The tagged autocorrelation separates into a fast part and a weak critical tail,
\begin{equation}
C_1(t)=C_{\rm fast}(\nu t)+N^{-1/2}C_{\rm crit}\!\left(\frac{\nu t}{\sqrt N}\right)+o(N^{-1/2}).
\label{eq:corrscale}
\end{equation}
The tail's shrinking weight compensates its growing lifetime. The integrated-noise asymptotics follow from count and tagged-contrast Poisson equations (Sec.~S2, ~\cite{sm}), rather than from pointwise scaling alone.

\paragraph{Time-average information lower bound.---}
For the stationary tagged time average
\begin{equation}
Y_T=\frac{1}{T}\int_0^T z_1(t)\,dt,
\end{equation}
$\mathrm{Var}(Y_T)=2T^{-1}\int_0^\infty C_1(t)dt+o(T^{-1})$. The score identity and Cauchy--Schwarz give the exact fixed-$N$ bound
\begin{equation}
R_{\rm loc}\ge R_{\rm mean}
\equiv\frac{\chi_1^2}{2\int_0^\infty C_1(t)\,dt}.
\label{eq:meanbound}
\end{equation}
For the stationary quartic diffusion $dx=-x^3du+\sqrt2\,dW_u$, define
\begin{equation}
\mathcal I=\int_0^\infty\!\langle x(u)x(0)\rangle_*\,du
=0.4874954944\ldots .
\end{equation}
The correlation area tends to $\kappa^{-1}(1+\mathcal I/c_s)$, yielding
\begin{equation}
\lim_{N\to\infty}\frac{R_{\rm mean}}{N\nu}
=C_s
=\frac{2(1-s)(1+s)^2}{c_s+\mathcal I}
\left[\frac{\Gamma(3/4)}{\Gamma(1/4)}\right]^2>0.
\label{eq:Cs}
\end{equation}

\paragraph{Local trajectory information rate.---}
Introduce the critical variables
\begin{equation}
\begin{aligned}
 x&=(Nc_s)^{1/4}m,
 &u&=\kappa\sqrt{\frac{c_s}{N}}\,t,\\
 \vartheta&=(1+s)c_s^{-1/4}N^{3/4}h.&&
\end{aligned}
\label{eq:scalingvars}
\end{equation}
The stationary signal converges to
\begin{equation}
 dx=(-x^3+\vartheta)du+\sqrt2\,dW_u.
\label{eq:signalSDE}
\end{equation}
Rescale the integrated tagged record as
\begin{equation}
Q_t=\int_0^t z_1(r)\,dr,
\qquad y_u=\kappa c_s^{1/4}N^{-1/4}Q_t.
\end{equation}
The tagged spin relaxes at rate $\kappa$ with quasistatic mean $m+o(N^{-1/4})$. Stochastic averaging gives the observation channel
\begin{equation}
 dy=\rho_s x\,du+\sqrt2\,dV_u,
\qquad\rho_s=c_s^{-1/2},
\label{eq:obsSDE}
\end{equation}
where $V$ and $W$ are independent. The martingale derivation and the asymptotic independence of the two noises are given in Sec. S2 of the Supplemental Material \cite{sm}. Define its stationary information rate for $\vartheta$ by
\begin{equation}
\mathcal J(\rho)=\lim_{U\to\infty}\frac1U
\mathbb E_0\!\left[
\left(\left.\partial_\vartheta\log p_\vartheta(y_{[0,U]})\right|_0\right)^2
\right].
\label{eq:Jdef}
\end{equation}
The change of field and clock is exact:
\begin{equation}
\left(\frac{d\vartheta}{dh}\right)^2\frac{du}{dt}
=N\kappa(1+s)^2.
\end{equation}

Transferring information requires control of likelihood scores, beyond process convergence~\cite{Catana2015,Surace2019}. Section~S2 (\cite{sm}) establishes convergence of both observation-only and hidden-endpoint information on each fixed critical-time interval, using the microscopic jump likelihood and uniform score-moment bounds. A uniform relaxation gap bounds the rate below by separated observed blocks and above by blocks with hidden boundary states supplied. Filter forgetting and reversibility make the endpoint information cost sublinear in block duration. The bounds converge to $\mathcal J(\rho_s)$ while preserving the stated order of limits. Hence the microscopic stationary rate obeys the proved asymptotic equality
\begin{equation}
R_{\rm loc}=N\nu(1-s)(1+s)^2\mathcal J(\rho_s)+o(N\nu).
\label{eq:RlocProved}
\end{equation}
For the complete site-resolved record, orthogonal compensated-jump scores and vanishing critical cavity fields give
\begin{equation}
R_{\rm full}=\frac{N\nu}{2}(1-s)(1+s)^2+o(N\nu).
\label{eq:Rfull}
\end{equation}
Equations~\eqref{eq:RlocProved} and \eqref{eq:Rfull} establish the limiting fraction in Eq.~\eqref{eq:central}.

The information rate $\mathcal{J}(\rho)$ is evaluated from the observation likelihood by nonlinear filtering. The filtering and parameter-sensitivity equations are given in Sec.~S2 of the Supplemental Material. The time-average and complete-score bounds imply $0<\eta_s^{\rm mean}\le\eta_s\le1$, with
\begin{equation}
\eta_s^{\rm mean}
=\frac4{c_s+\mathcal I}
\left[\frac{\Gamma(3/4)}{\Gamma(1/4)}\right]^2
=\frac{V_x^2}{c_s+\mathcal I}.
\label{eq:etamean}
\end{equation}

The lower bound is minimized at $s=1/3$, where $c_s=3/8$, yielding

\begin{equation}
\eta_s \ge \eta_s^{\mathrm{mean}}
\ge \eta_{1/3}^{\mathrm{mean}}
= 0.5297959050\ldots > \frac{1}{2},
\quad 0 \le s < 1 
\end{equation}

\begin{figure}[!tb]
\centering
\includegraphics[width=\columnwidth]{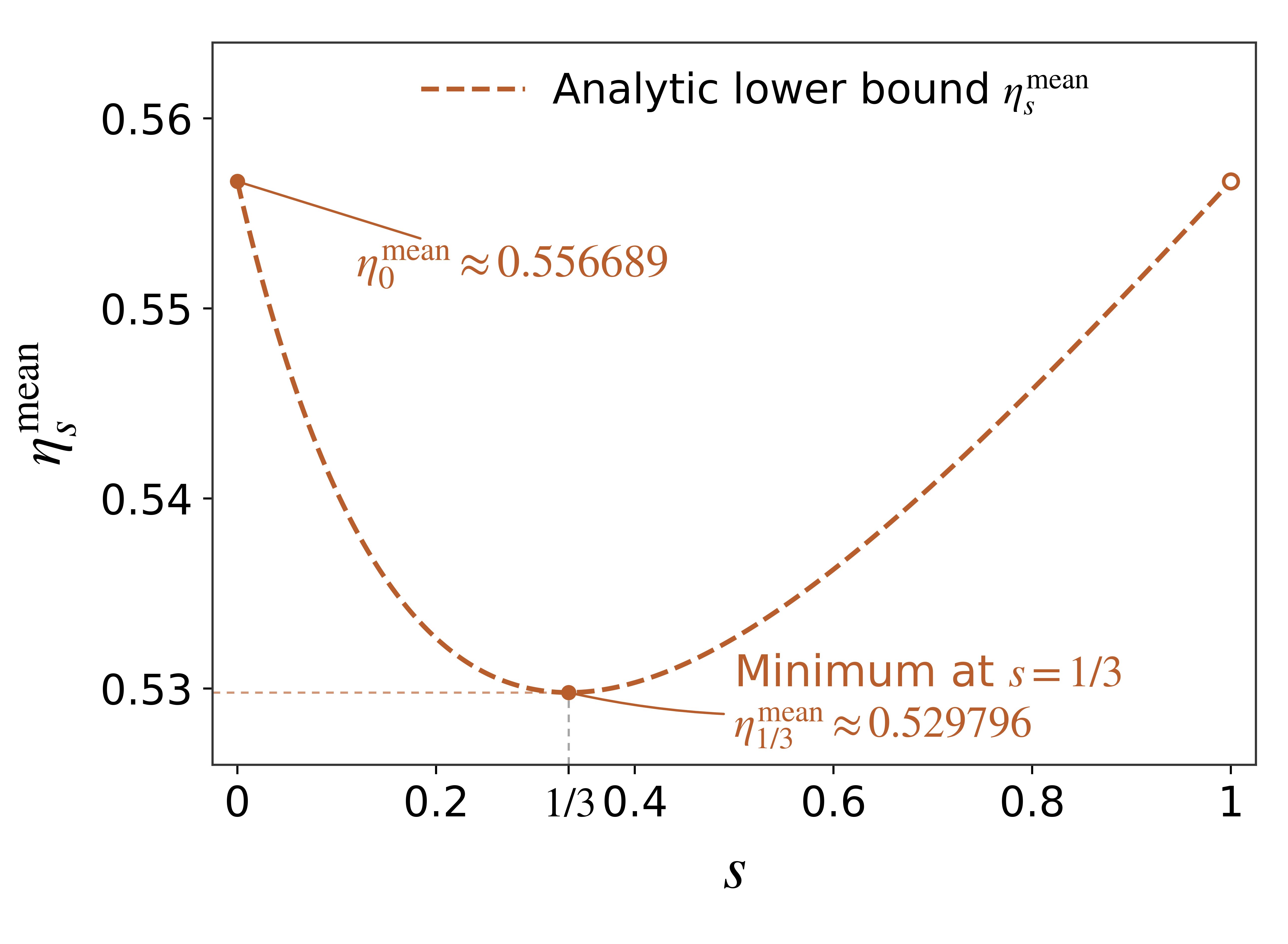}
\caption{Analytic time-average lower bound $\eta_s^{\mathrm{mean}}$
on local information retention versus $s$.
The dashed curve is evaluated at $h=0$ and $J=J_c=1/(1+s)$ in the rapid-readout critical regime. The minimum occurs at $s=1/3$, where
$\eta_{1/3}^{\mathrm{mean}}\approx 0.529796$. At $s=0$, $\eta_0^{\mathrm{mean}}\approx 0.556689$; the same value is approached as $s\to 1^{-}$, as indicated by the open circle.}
\label{fig:sim}
\end{figure}

This more-than-half guarantee is analytic and does not rely on numerical filtering. Fig~\ref{fig:sim} illustrates the analytic lower bound $\eta_s^{\mathrm{mean}}$ as a function
of $s$.

At $s=1/3$, numerical filtering gives
$\eta_{1/3}\simeq 0.532$, close to the analytic lower bound. The small difference between the numerical estimate and the mean bound does not establish strict additional information
at the quoted precision.

The joint time average of $k$ yields the following lower bound on the retained Fisher-information fraction:
\begin{equation} \eta_{k,s}^{\mathrm{mean}}
    = \frac{kV_x^2}{c_s+k\mathcal I}.
\label{eq:mean}
\end{equation}

\begin{equation}
\lim_{k \to \infty} \eta_{k,s}^{\mathrm{mean}}
= \frac{V_x^2}{\mathcal I}
\approx 0.937335.
\end{equation}

\paragraph{Sparse critical readout.---}
A growing set of local trajectories permits asymptotically complete
readout of the collective information.(Sec. S4, \cite{sm}) Let $R_{N,k}$ be the stationary information rate
in the complete trajectories of any $k$ distinct nodes, with the same
common-field, critical, and rapid-readout assumptions as above; thus
$R_{N,1}=R_{\rm loc}$ and $R_{N,N}=R_{\rm full}$. For each fixed $k$,
\begin{equation}
\eta_{k,s}\equiv\lim_{N\to\infty}\frac{R_{N,k}}{R_{N,N}}
=2\mathcal J(\sqrt{k}\,\rho_s).
\label{eq:sparse_fraction}
\end{equation}
The $k$ outputs share one collective signal, while their fast noises
become independent. Their normalized sum therefore increases the
observation strength by $\sqrt{k}$, rather than supplying independent
copies of the collective information.

For fixed $s$, the \emph{sparse critical readout theorem} gives

\begin{equation}
 \frac{9c_sV_x^2}{k+9c_s}
 \ \le\ 1-\eta_{k,s}
 \ \le\ \frac{9c_s}{k-6c_s},\qquad k>6c_s.
 \label{eqS:sparse_two_sided}
\end{equation}

\begin{figure}[!tb]
\centering
\includegraphics[width=\columnwidth]{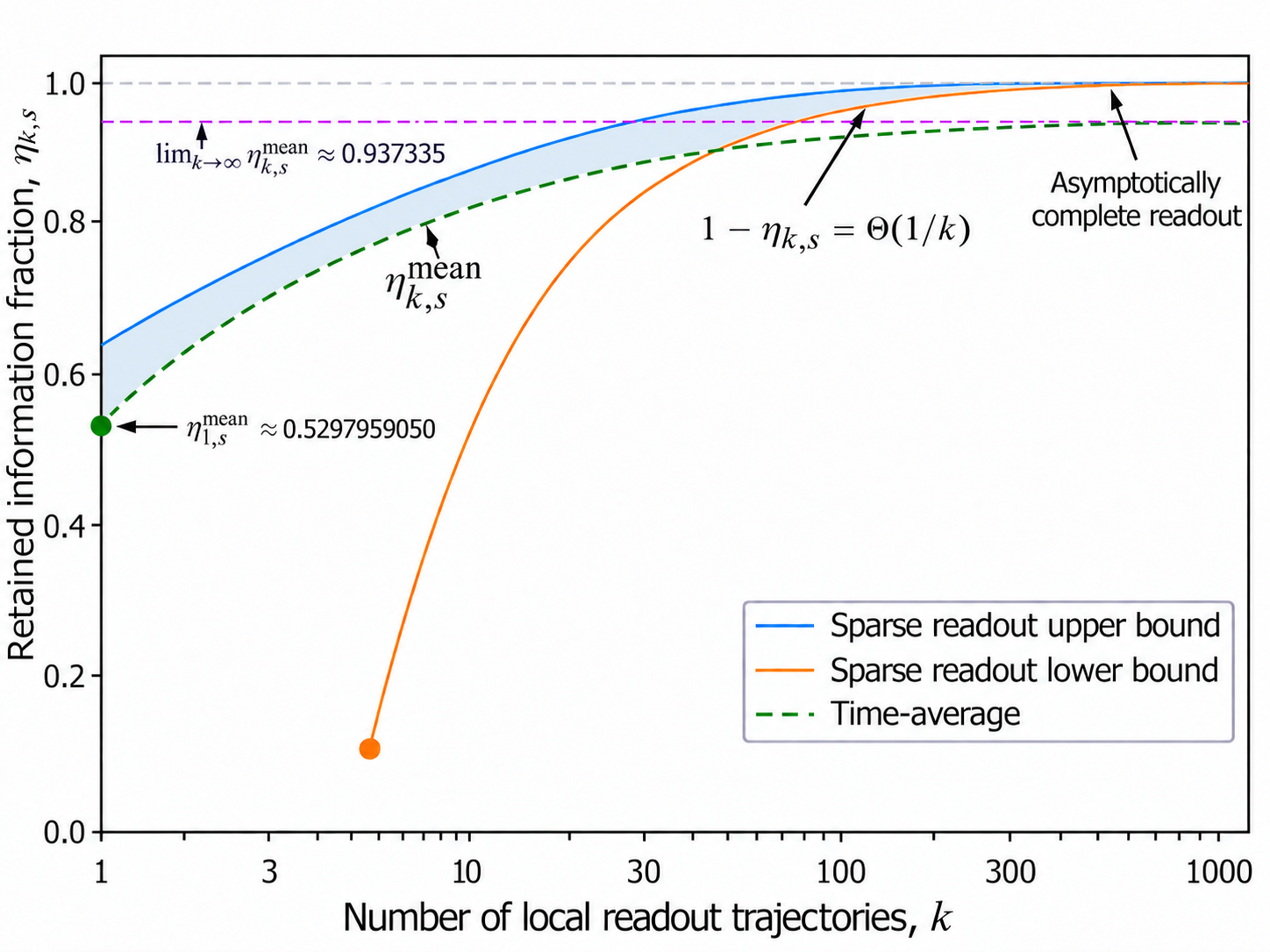}
\caption{Sparse critical readout at $s=1/3$. Retained Fisher-information fraction $\eta_{k,s}$ as a function of the number $k$ of local readout trajectories. The shaded region represents the analytically allowed
range of $\eta_{k,s}$, bounded above by the sparse-readout
upper bound and below by the tighter of the time-average and sparse-readout lower bounds.
The lower boundary switches from the time-average bound to the sparse-readout bound at $k\simeq 45$.
}
\label{fig:sim1}
\end{figure}

These bounds establish the asymptotic information-loss
scaling $1-\eta_{k,s}=\Theta(k^{-1})$ as $k\to\infty$
and the readout cost
$k_{\min}(\varepsilon;s)=\Theta(\varepsilon^{-1})$
as $\varepsilon\downarrow 0$, where
$k_{\min}(\varepsilon;s)
=\min\{k\in\mathbb{N}:\eta_{k,s}\geq 1-\varepsilon\}$. Thus, the number of readout trajectories required to retain a prescribed asymptotic information fraction is independent of the network size. Fig.~\ref{fig:sim1} illustrates these analytic
bounds for $s=1/3$, where $c_s=3/8$, together with the time-average lower bound in Eq.~\eqref{eq:mean}.

Moreover, any growing readout set of vanishing density satisfies
\begin{equation}
\begin{gathered}
k_N\to\infty,\qquad k_N/N\to0\quad\Longrightarrow\\[-2pt]
\frac{R_{N,k_N}}{R_{N,N}}\longrightarrow1.
\end{gathered}
\label{eq:sparse_complete}
\end{equation}

This follows from monotonicity in the number of retained trajectories
and the fixed-$k$ limits, without assuming a uniform finite-$N$
$1/k$ bound. The completeness concerns the common-field information
rate in temporal records; the controller still uses network-wide
measurements.

\paragraph{Parameter directions and general networks.—}
Detailed derivations of the results below are given in Sec. S3 of the Supplemental Material \cite{sm}.

Criticality alone does not guarantee that collective information is visible in a given local record: the parameter must also drive a direction that overlaps with the locally accessible collective response.
To make this explicit, replace the common field by spatially varying fields
$h_j=\theta g_j$, with $\|g\|^2=\Theta(N)$ and $g_1=O(1)$.
Permutation symmetry projects the local score onto the uniform component of the drive, while the tagged-site contrast carries only a vanishing fraction of the full information budget.
If
\[
\Gamma_N
=
\frac{N\bar g^{\,2}}{\|g\|^2}
\longrightarrow \Gamma,
\qquad
\bar g=\frac{1}{N}\sum_j g_j,
\]
then
\begin{equation}
\frac{R_{\rm loc}(g)}{R_{\rm full}(g)}
\longrightarrow
\Gamma\,\eta_s .
\label{eq:number1}
\end{equation}
Thus a uniform drive acting on a fraction $p$ of the sites retains
$p\eta_s$, even when the observed site lies outside the driven set,
whereas a balanced drive with vanishing spatial mean has a vanishing
asymptotic retained fraction.

This visibility principle extends beyond all-to-all symmetry.
Consider the stable Gaussian network
$dZ=-\kappa A Z\,dt+\kappa b\theta\,dt+\sqrt{2\kappa D}\,dW$,
with $D>0$, and a scalar readout $Y=c^{T}Z$.
The exact retained fraction is
\begin{equation}
\frac{R_Y}{R_{\rm full}}
=
\frac{(c^{T}A^{-1}b)^2}
{\bigl(c^{T}A^{-1}DA^{-T}c\bigr)
 \bigl(b^{T}D^{-1}b\bigr)} .
\label{eq:number2}
\end{equation}
Equation~\eqref{eq:number2} is the squared overlap between the noise-weighted drive
direction $D^{-1/2}b$ and the readout-response direction
$D^{1/2}A^{-T}c$.
Hence a slow collective mode enhances locally accessible information
only when it has finite weight at the readout and is aligned with the
driven parameter direction.
Together with Eqs.~\eqref{eq:RlocProved} and ~\eqref{eq:Rfull}, these results separate two ingredients:
criticality sets the extensive information scale, whereas mode alignment
determines how much of that information is visible in a local trajectory.

\paragraph{Conclusion.---}
We have established a microscopic trajectory-rate theorem for sensing from a restricted readout. Although the controller uses the network-wide record, the estimator retains only one tagged trajectory. In the rapid-readout common-field critical regime, this single binary trajectory inherits the extensive scaling of the full information rate despite critical slowing. The growing critical lifetime is compensated by the shrinking local weight of the slow mode, leaving finite integrated local noise. The analytic lower bound remains above one half throughout the channel-parameter range and is minimized at $s=1/3$; there the complete likelihood gives a retained fraction of approximately $53\%$. Remarkably, this information concentration extends to sparse readout:
the fractional information loss scales as
$1-\eta_{k,s}=\Theta(k^{-1})$, and any growing set of
retained trajectories satisfying $k_N\to\infty$ and
$k_N/N\to 0$ asymptotically recovers the full-network
Fisher-information rate for the common weak field.
Thus, criticality enables asymptotically complete access
to collective sensing information through a vanishing
fraction of local readout records.

Beyond the all-to-all setting, our results separate two ingredients of local access to collective information: criticality sets the information scale, whereas the overlap between the parameter-driven collective response and the readout controls its local visibility. The Gaussian-network and nonlinear-observation extensions show that this visibility principle is not tied to permutation symmetry or to mean response alone. These results connect naturally to distributed sensing and partially observed dynamical networks, where collective information must be inferred from restricted readouts. An important next step is to determine whether the same critical information-concentration mechanism persists in finite-connectivity nonlinear networks, and at finite measurement cadence and observation time.

\begin{acknowledgments}
This work is supported by the Key Research and Development Support Program of the Chengdu Municipal Science and Technology Bureau through the project ``Research, Development, and Application Demonstration Platform for an Independently Controllable Room-Temperature Quantum--Supercomputing Integrated System'' (Grant No. 2025-XT00-00012-GX). KLC acknowledges funding from the National Research Foundation, Singapore and the Ministry of Education, Singapore.
\end{acknowledgments}

\bibliography{references}
\end{document}


\title{Supplemental Material for ``Criticality Concentrates Extensive Sensing Information into a Local Readout''}


\author{Peng Wang}
\email{wp002005@163.com}
\altaffiliation{These authors contributed equally to this work.}
\affiliation{School of Computer and Artificial Intelligence, Southwest Minzu University, Chengdu 610225, China}
\affiliation{Centre for Quantum Technologies, National University of Singapore, Singapore 117543, Singapore}

\author{Yu-Xuan Zhang}
\email{2411915@mail.nankai.edu.cn}
\altaffiliation{These authors contributed equally to this work.}
\affiliation{School of Physics, Nankai University, Tianjin 300071, China}
\affiliation{Centre for Quantum Technologies, National University of Singapore, Singapore 117543, Singapore}

\author{Hai-Tao Ding}
\email{htding.9@nus.edu.sg}
\affiliation{Centre for Quantum Technologies, National University of Singapore, Singapore 117543, Singapore}
\affiliation{MajuLab, CNRS-UNS-NUS-NTU International Joint Research Unit, Singapore UMI 3654, Singapore}

\author{Leong-Chuan Kwek}
\email{kwekleongchuan@nus.edu.sg}
\affiliation{Centre for Quantum Technologies, National University of Singapore, Singapore 117543, Singapore}
\affiliation{MajuLab, CNRS-UNS-NUS-NTU International Joint Research Unit, Singapore UMI 3654, Singapore}
\affiliation{National Institute of Education, Nanyang Technological University, Singapore 637616, Singapore}

\pagenumbering{roman}
\maketitle
\thispagestyle{empty}

\begingroup
  \setcounter{tocdepth}{1}

  \makeatletter
  \let\l@subsection\@gobbletwo
  \let\l@subsubsection\@gobbletwo
  \makeatother

  \hypersetup{linkcolor=red}
  \tableofcontents
\endgroup

\clearpage
\pagenumbering{arabic}
\setcounter{page}{0}

\section{Spin dynamics and critical scaling}

\subsection{Spin process}

In the rapid-readout process of the Letter, the state is specified by
$N$ spins $z_i=\pm1$. The common field $h$ is the parameter to be
estimated, $J$ is the coupling, and $\nu>0$ and $0\le s<1$ fix the
flip dynamics. For $N\ge2$, the field and flip rate at site $i$ are
\begin{align}
 a_i(\bm z;h)&=h+\frac{J}{N-1}\sum_{j\ne i}z_j,\\
 w_i(\bm z;h)&=\frac\kappa2
 [1-z_i\tanh a_i][1-sz_i\tanh a_i],\qquad
 \kappa=\nu(1-s).
 \label{eqS:rate}
\end{align}
Write $w_z(a)$ for this rate at sign $z$ and field $a$.
The local record contains only $z_1(t)$; the complete record contains
all spin paths. We estimate $h$ at $h=0$ and at the critical coupling
derived below. For a stationary record of length $T$, let
$F_h^{\rm loc}(T)$ and $F_h^{\rm full}(T)$ be their Fisher information.
The corresponding rates are
\[
 R_{\rm loc}=\lim_{T\to\infty}\frac{F_h^{\rm loc}(T)}T,
 \qquad
 R_{\rm full}=\lim_{T\to\infty}\frac{F_h^{\rm full}(T)}T.
\]
The time limit is taken at each fixed $N$, followed by $N\to\infty$,
with $s$ and $\nu$ fixed. At criticality, the response shared by the
spins allows one tagged trajectory to retain a nonzero fraction of
the complete-record information. Its susceptibility grows as
$\sqrt N$, while its integrated noise remains finite: the collective
fluctuation lasts longer but has a smaller weight in a single-spin
correlation. The critical dynamics below fixes these two scales.

At a fixed count $K$ of positive spins, all negative spins have the
same upward flip rate and all positive spins have the same downward
flip rate. Hence $K$ is itself a birth--death chain. With
$m_K=2K/N-1$, its rates are
\begin{equation}
 \begin{aligned}
 \Lambda_K&=(N-K)w_-\left(h+J\frac{2K-N+1}{N-1}\right),\\
 M_K&=Kw_+\left(h+J\frac{2K-N-1}{N-1}\right).
 \end{aligned}
 \label{eq:count-rates}
\end{equation}
The pair $(z_1,K)$ is also Markov; write $Q_h$ for its generator.
Keeping the tagged sign will allow local correlations and likelihoods
to be calculated without retaining the other individual spin labels.

\subsection{Critical point and stationary distribution}

The magnetization $m=N^{-1}\sum_i z_i$ changes by
$\Delta_i m=-2z_i/N$. Set $b=\tanh(h+Jm)$.
The cavity correction is $O(N^{-1})$, so
\[
 w_i=\frac\kappa2[1-(1+s)z_i b+sb^2]+O(\kappa/N).
\]
Its drift and quadratic variation follow by summing these jumps:
\begin{align}
 A(m)&=\sum_i\Delta_i m\,w_i
 =\frac2N(\Lambda_K-M_K)\nonumber\\
 &=\kappa[(1+s)b-(1+sb^2)m]+O(\kappa/N),
 \label{eqS:drift_exact}\\
 B(m)&=\sum_i(\Delta_i m)^2w_i
 =\frac4{N^2}(\Lambda_K+M_K)\nonumber\\
 &=\frac{2\kappa}{N}[1-(1+s)bm+sb^2]+O(\kappa/N^2).
\end{align}
In particular,
\begin{equation}
 B(m)=\frac{2\kappa}{N}+o(N^{-1})\qquad(m,h\to0).
 \label{eqS:diffm}
\end{equation}
Expanding $b=(h+Jm)-(h+Jm)^3/3+O((h+Jm)^5)$ gives
\[
 A_{\rm lin}(m)=\kappa(1+s)h-\kappa[1-(1+s)J]m.
\]
The linear restoring term vanishes at
\begin{equation}
 J_c=\frac1{1+s}.
 \label{eqS:Jc}
\end{equation}
At this coupling, the cubic coefficient is
\begin{equation}
 c_s=\frac{1+s}{3}J_c^3+sJ_c^2
     =\frac{s+1/3}{(1+s)^2}>0.
 \label{eqS:cs}
\end{equation}
The leading diffusion equation near the critical point is therefore
\begin{equation}
 dm=[\kappa(1+s)h-\kappa c_sm^3
       +O(\kappa m^5,\kappa hm^2,\kappa h^3,\kappa/N)]dt
       +\sqrt{2\kappa/N}\,dW_t.
 \label{eqS:critical_SDE_m}
\end{equation}

For the edge $K\leftrightarrow K+1$, the field entering the upward
and downward rates is the same. Their ratio is
$w_-(a)/w_+(a)=e^{2F_s(a)}$, where
$F_s(a)=a+\operatorname{arctanh}(s\tanh a)$. The stationary count
weights therefore satisfy
\begin{align}
 P_K\Lambda_K&=P_{K+1}M_{K+1},\nonumber\\
 \log\frac{P_{K+1}}{P_K}
 &=\log\frac{N-K}{K+1}
   +2F_s\left(h+J\frac{2K-N+1}{N-1}\right).
 \label{eq:count-weights}
\end{align}
Each configuration of count $K$ has weight $P_K/\binom NK$.
The same equality balances every spin-flip edge, proving reversibility
of the spin chain and its $(z_1,K)$ reduction.

Near zero,
\[
 \begin{aligned}
 F_s(a)&=(1+s)a-\frac{s(1-s^2)}3a^3+O(a^5),\\
 \log\frac{1-m}{1+m}&=-2m-\frac23m^3+O(m^5).
 \end{aligned}
\]
At $J_c$, the linear terms cancel and the cubic coefficient is
$-2c_s$. Summing the ratios with spacing $2/N$ gives
\begin{equation}
 P_N(m)\propto\exp[-Nc_sm^4/4+o(1)],\qquad h=0.
 \label{eqS:quartic_density}
\end{equation}
Thus
\begin{equation}
 m=O(N^{-1/4}).
 \label{eqS:m_scale}
\end{equation}
For a small field, the leading exponent is
\begin{equation}
 -Nc_sm^4/4+N(1+s)hm.
 \label{eqS:field_weight}
\end{equation}

\subsection{Critical scaling}

The terms compete when $h=O(N^{-3/4})$. Define the critical scales by
\begin{equation}
 \begin{aligned}
 m&=\epsilon_Nx,&u&=\omega_Nt,&\vartheta&=\alpha_Nh,\\
 \epsilon_N&=(Nc_s)^{-1/4},&
 \omega_N&=\kappa\sqrt{c_s/N},&
 \alpha_N&=(1+s)c_s^{-1/4}N^{3/4}.
 \end{aligned}
 \label{eqS:critical_scaling}
\end{equation}
At $m=\epsilon_Nx$, the cubic drift has size
$\kappa c_s\epsilon_N^3$. Dividing by the amplitude $\epsilon_N$
gives the relaxation rate $\omega_N=\kappa c_s\epsilon_N^2$.
The field term and quadratic variation then scale as
\[
 \frac{\kappa c_s\epsilon_N^2}{\omega_N}=1,\qquad
 \frac{\kappa(1+s)}{\alpha_N\epsilon_N\omega_N}=1,\qquad
 \frac{2\kappa}{N\epsilon_N^2\omega_N}=2.
\]
These balances give the critical process
\begin{equation}
 dx=(-x^3+\vartheta)du+\sqrt2\,dW_u.
 \label{eqS:signal_final}
\end{equation}
Thus the physical critical lifetime is
$\omega_N^{-1}=O(\sqrt N/\nu)$, and the field window is
\begin{equation}
 h=\frac{c_s^{1/4}}{(1+s)N^{3/4}}\,\vartheta.
 \label{eqS:h_theta}
\end{equation}

The zero-field stationary density of this process is
\begin{equation}
 p_*(x)=Z^{-1}e^{-x^4/4},\qquad
 Z=\int_{-\infty}^\infty e^{-x^4/4}dx,
 \label{eqS:pstar}
\end{equation}
and substitution of $v=x^4/4$ gives its variance,
\begin{equation}
 V_x\equiv\langle x^2\rangle_*
 =2\frac{\Gamma(3/4)}{\Gamma(1/4)}.
 \label{eqS:Vx}
\end{equation}

The stationary weights of the finite chain approach this density
on the same scale. This also controls the moments used in the response
and noise calculations. The count spacing in $x$ is $d_N=2/(N\epsilon_N)$.
On bounded $x_K=m_K/\epsilon_N$ and $\vartheta$ intervals, the exact
ratio sum and its parameter derivative give
\begin{equation}
 \log P_K=C_N-x_K^4/4+\vartheta x_K+O(N^{-1/2}).
 \label{eq:stationary-control}
\end{equation}
To control the tails, put $t_a=\tanh a$. For $0<m<1$,
\[
 J_cF_s'(J_cm)
 =\frac1{1+s}\left[1+s\frac{1-t_{J_cm}^2}{1-s^2t_{J_cm}^2}\right]
 \le1<\frac{d}{dm}\operatorname{arctanh}m.
\]
Hence $\operatorname{arctanh}m-F_s(J_cm)\ge cm^3$:
its expansion is $c_sm^3+O(m^5)$ near zero, and its ratio to $m^3$
stays positive away from zero. Summing the decreasing count ratios yields
\begin{equation}
 P_K\le Cd_Ne^{-cx_K^4+C|x_K|},\qquad
 \sup_N\mathbb E_\vartheta|x_K|^r<\infty\quad(r\ge0\text{ fixed}).
 \label{eq:stationary-tails}
\end{equation}

The bound permits stationary averages and their field derivatives
to pass to the limit. For the dynamics, Taylor expansion of the count generator therefore gives, on $|x|\le M$,
\[
 \omega_N^{-1}\mathcal G_Nf(x)
 =[\vartheta-x^3+O_M(N^{-1/4})]f'(x)
  +[1+O_M(N^{-1/2})]f''(x)+O_M(d_N\|f^{(3)}\|_\infty).
\]
The vanishing jumps, generator limit, and stationary moment bound
establish Eq.~\eqref{eqS:signal_final} on every fixed critical-time
interval, starting from the stationary state.

\section{Local Fisher information}

The stationary information rates at $J=J_c$ and $h=0$ are
\[
 R_{\rm loc}=N\kappa(1+s)^2\mathcal J(\rho_s)+o(N\nu),
 \qquad
 R_{\rm full}=\frac{N\kappa}{2}(1+s)^2+o(N\nu),
 \qquad \rho_s=c_s^{-1/2}.
\]
Here $\mathcal J(\rho)$ is the information rate for $\vartheta$ in
the noisy observation $dy=\rho x\,du+\sqrt2\,dV_u$ of the critical
signal \eqref{eqS:signal_final}, with $V$ independent of its driving
noise. Thus one spin retains the fraction $2\mathcal J(\rho_s)$.
The time average already gives a positive lower bound on this
fraction. The likelihood of the whole tagged path determines the
coefficient $\mathcal J$.

\subsection{Response and integrated noise}

For a statistic $X_T$ of the tagged record, let the record score be
$S_T=\partial_h\log P_h^T|_0$. Differentiating its mean gives
$\partial_h\mathbb E_hX_T|_0=\mathbb E_0[(X_T-\mathbb E_0X_T)S_T]$.
Cauchy--Schwarz then yields
\begin{equation}
 F_h^{\rm loc}(T)=\mathbb E_0S_T^2
 \ge\frac{[\partial_h\mathbb E_hX_T|_0]^2}
           {\operatorname{Var}_0X_T}.
 \label{eqS:CS_FI}
\end{equation}
For the time average $\bar z_T=T^{-1}\int_0^Tz_1(t)dt$, stationarity
gives
\[
 \left.\partial_h\mathbb E_h\bar z_T\right|_0=\chi_1,
 \qquad \operatorname{Var}_0\bar z_T
 =\frac2{T^2}\int_0^T(T-t)C_1(t)dt,
\]
where $\chi_1=\partial_h\langle z_1\rangle_h|_0$ and
$C_1(t)=\langle z_1(t)z_1(0)\rangle_0$. The zero-field mean is zero.
At fixed $N$, the finite chain mixes exponentially, so
$T\operatorname{Var}_0\bar z_T\to2\int_0^\infty C_1(t)dt$.
The information bound becomes
\begin{equation}
 R_{\rm loc}\ge R_{\rm mean}
 =\frac{\chi_1^2}{2\int_0^\infty C_1(t)dt}.
 \label{eqS:Rmean_bound}
\end{equation}

Exchangeability gives $\langle z_1\rangle_h=\langle m\rangle_h$.
At zero field, let $r_N(K)=\partial_h\log P_K|_0$ be the stationary
score. Symmetry and differentiation of Eq.~\eqref{eq:count-weights} give
\[
 \begin{aligned}
 r_N(N-K)&=-r_N(K),\\
 r_N(K+1)-r_N(K)
 &=2F_s'\left(J_c\frac{2K-N+1}{N-1}\right).
 \end{aligned}
\]
Summing from the central count and using $F_s'(a)=1+s+O(a^2)$ yields
\[
 r_N(K)=N(1+s)m_K+O(N|m_K|^3+|m_K|).
\]
The local susceptibility is consequently
\begin{align}
 \chi_1&\equiv\left.\partial_h\langle z_1\rangle_h\right|_0
 =\mathbb E_0[m_Kr_N(K)]\nonumber\\
 &=N(1+s)\operatorname{Var}_0(m)+O(1),
 \label{eqS:chi_fluct}
\end{align}
because the quartic moment bound gives $N\mathbb E_0m^4=O(1)$.
Equations~\eqref{eq:stationary-control} and \eqref{eqS:Vx} give
\[
 \operatorname{Var}_0(m)=\frac{V_x}{\sqrt{Nc_s}}+O(N^{-1}).
\]
Substitution in Eq.~\eqref{eqS:chi_fluct} gives
\begin{equation}
 \chi_1=A_s\sqrt N+O(1),\qquad
 A_s=\frac{2(1+s)}{\sqrt{c_s}}
       \frac{\Gamma(3/4)}{\Gamma(1/4)}.
 \label{eqS:chi_final}
\end{equation}

The numerator of Eq.~\eqref{eqS:Rmean_bound} is therefore of order
$N$. Its denominator includes both collective magnetization
fluctuations and fluctuations of the tagged spin at a fixed count.
These contributions separate under
\[
z_1=m+\delta_1,
 \qquad
 \delta_1\equiv z_1-m.
\]
Permutation symmetry gives
\[
\langle z_1(t)m(0)\rangle
 =\frac1N\sum_j\langle z_j(t)m(0)\rangle
 =\langle m(t)m(0)\rangle,
\]
and similarly $\langle m(t)z_1(0)\rangle=\langle m(t)m(0)\rangle$.  The cross terms therefore cancel, leaving
\begin{equation}
 C_1(t)\equiv\langle z_1(t)z_1(0)\rangle
 =\langle m(t)m(0)\rangle
 +\langle\delta_1(t)\delta_1(0)\rangle.
 \label{eqS:corr_decomp}
\end{equation}
The two components have different amplitudes and lifetimes.
The collective variance is $O(N^{-1/2})$, but it persists for a
time $O(\omega_N^{-1})$; their product is $O(\kappa^{-1})$.
The residual tagged fluctuation has order-one variance and relaxes
on the microscopic time scale. Their respective correlation areas
are $\mathcal I/(\kappa c_s)$ and $1/\kappa$ to leading order, where
\begin{equation}
 \mathcal I=\int_0^\infty\langle x(u)x(0)\rangle_*\,du.
 \label{eqS:Idef}
\end{equation}
Their sum is therefore
\begin{equation}
 \int_0^\infty C_1(t)dt
 =\frac1\kappa\left(1+\frac{\mathcal I}{c_s}\right)
 +o(\kappa^{-1}).
 \label{eqS:total_area}
\end{equation}
Substitution of the response and noise area yields
\begin{equation}
 \lim_{N\to\infty}\frac{R_{\rm mean}}{N\nu}
 =C_s
 =\frac{2(1-s)(1+s)^2}{c_s+\mathcal I}
   \left[\frac{\Gamma(3/4)}{\Gamma(1/4)}\right]^2>0.
 \label{eqS:Cs_final}
\end{equation}
The growing response and finite noise area give an extensive
information rate using only $\bar z_T$. The following count and
contrast calculations determine the two areas, including their
large-$N$ limits.

For the collective component, integrating the stationary correlation
is equivalent to solving a Poisson equation for the count generator
$\mathcal G_N$. If $g$ satisfies
\begin{equation*}
 -\mathcal G_Ng=m,\qquad \langle g\rangle_P=0,
\end{equation*}
then its area is $\langle m,g\rangle_P$. In a birth--death chain,
this solution is obtained by summing across consecutive edges.
Define
\begin{equation*}
 c_K=P_K\Lambda_K,\qquad b_K=\sum_{j=0}^KP_jm_j.
\end{equation*}
Detailed balance turns the Poisson equation into an edge difference.
Its partial sum fixes each edge gradient:
\[
 \begin{aligned}
 -P_Km_K
 &=c_K(g_{K+1}-g_K)-c_{K-1}(g_K-g_{K-1}),\\
 c_K(g_{K+1}-g_K)&=-\sum_{j=0}^KP_jm_j=-b_K.
 \end{aligned}
\]
The stationary correlation area is consequently
\begin{align}
 A_m&\equiv\int_0^\infty\langle m(t)m(0)\rangle\,dt
     =\langle m,g\rangle_P\nonumber\\
 &=\sum_{K=0}^{N-1}c_K(g_{K+1}-g_K)^2
  =\sum_{K=0}^{N-1}\frac{b_K^2}{P_K\Lambda_K}.
 \label{eq:count-area}
\end{align}

In the critical coordinates,
\begin{equation*}
 \frac{P_K}{d_N}\longrightarrow p_*(x_K),\qquad
 \Lambda_K\sim\frac{N\kappa}{4},\qquad
 \frac{b_K}{\epsilon_N}\longrightarrow
      \int_{-\infty}^{x_K}yp_*(y)\,dy.
\end{equation*}

For the tails, $\mathbb E_0m=0$ gives
\begin{equation*}
 b_K=-\sum_{j>K}P_jm_j,\qquad
 |b_K|\le\frac{CP_K}{m_K^2}
 \quad(m_K> M\epsilon_N).
\end{equation*}
The bound follows by summing the decreasing count ratios; the negative
tail follows by symmetry. Together with Eq.~\eqref{eq:stationary-tails},
it gives
\begin{equation*}
 \limsup_{N\to\infty}\kappa
 \sum_{|x_K|>M}\frac{b_K^2}{P_K\Lambda_K}
 \le C\int_{|x|>M}e^{-c x^4}\,dx\longrightarrow0.
\end{equation*}

The full sum can now be passed to the limit:
\begin{align}
 A_m
 &=\frac{1}{\kappa c_s}
    \int_{-\infty}^{\infty}
       \frac{[\int_{-\infty}^xyp_*(y)\,dy]^2}{p_*(x)}\,dx
       +o(\kappa^{-1})\nonumber\\
 &=\frac{\mathcal I}{\kappa c_s}+o(\kappa^{-1}).
 \label{eqS:slow_area}
\end{align}

The remaining term in Eq.~\eqref{eqS:corr_decomp} is the correlation
of $\delta_1=z_1-m$. At $J=J_c$, the tagged cavity field is
\begin{align}
 a_1
 &=h+\frac{J_c}{N-1}(Nm-z_1)\nonumber\\
 &=\frac{m}{1+s}+o(N^{-1/4})
 \label{eqS:a1_expand}
\end{align}
in the critical window $m=O(N^{-1/4})$, $h=O(N^{-3/4})$.  Expanding Eq.~\eqref{eqS:rate} then gives, for a tagged state $z=\pm1$,
\begin{equation}
 w_z=\frac{\kappa}{2}(1-zm)+O(N^{-1/2}).
 \label{eqS:telegraph_rate}
\end{equation}
Conditional on the count, the tagged sign satisfies
\begin{equation*}
 \mathbb E[z_1\mid K]=m_K,\qquad
 \mathbb E[(z_1-m_K)^2\mid K]=1-m_K^2.
\end{equation*}
A function of $(z_1,K)$ therefore splits into a count function and
$(z_1-m_K)a_K$, orthogonally in stationarity. The second component
describes which spin differs from the population mean at a given
count. Its evolution allows the count to move, but also damps that
tagged-site contrast. Applying $Q_0$ gives
$Q_0[(z_1-m_K)a_K]=(z_1-m_K)(\mathcal G_\delta a)_K$, with
\begin{equation}
 (\mathcal G_\delta a)_K
 =\widetilde\Lambda_K(a_{K+1}-a_K)
  +\widetilde M_K(a_{K-1}-a_K)-\ell_Ka_K.
 \label{eq:contrast-generator}
\end{equation}
The modified count rates and contrast decay rate are
\begin{equation*}
 \begin{aligned}
 \widetilde\Lambda_K&=(N-K-1)w_-(J_cg_-),&
 \widetilde M_K&=(K-1)w_+(J_cg_+),\\
 g_\pm&=\frac{2K-N\mp1}{N-1},&
 \ell_K&=w_-(J_cg_-)+w_+(J_cg_+).
 \end{aligned}
\end{equation*}
Here $1\le K\le N-1$. The bounded cavity fields give
\begin{equation*}
 \ell_K\ge c\kappa>0.
\end{equation*}
This positive decay rate keeps the contrast on the microscopic
time scale even when the count itself relaxes critically.
Let $K_r$ be the count chain with rates
$\widetilde\Lambda_K,\widetilde M_K$. Conditioning on its first jump
solves $-\mathcal G_\delta a=1$ as
\begin{equation*}
 a_K=\mathbb E_K\int_0^\infty
          e^{-\int_0^t\ell_{K_r}\,dr}\,dt,
 \qquad 0\le a_K\le\frac1{c\kappa}.
\end{equation*}
The stationary Poisson formula for $\delta_1$ is therefore
\begin{equation}
 A_\delta=\sum_{K=1}^{N-1}P_K(1-m_K^2)a_K.
 \label{eq:contrast-area}
\end{equation}
On a fixed physical-time interval, a stationary initial count stays
in the critical window with probability tending to one. There
$m_K\to0$ and $\ell_K\to\kappa$, while
\begin{equation*}
 0\le e^{-\int_0^t\ell_{K_r}dr}\le e^{-c\kappa t}.
\end{equation*}
Dominated convergence in Eq.~\eqref{eq:contrast-area} gives
\begin{equation}
 A_\delta=\int_0^\infty\langle\delta_1(t)\delta_1(0)\rangle\,dt
 =\frac1\kappa+o(\kappa^{-1}).
 \label{eqS:fast_area}
\end{equation}

The constant $\mathcal I$ fixes the coefficient in
Eq.~\eqref{eqS:Cs_final}. For the limiting generator $\mathcal Lf=-x^3f'+f''$, the same
identity gives $\mathcal I=\langle x,(-\mathcal L)^{-1}x\rangle_*$.
Writing $\phi=(-\mathcal L)^{-1}x$ reduces its evaluation to
\begin{equation}
 -\mathcal L\phi=x.
 \label{eqS:Poisson_phi}
\end{equation}
Because $p_*(x)\propto e^{-x^4/4}$, Eq.~\eqref{eqS:Poisson_phi} can be written as
\[
-\frac{d}{dx}\left(e^{-x^4/4}\phi'(x)\right)
 =x e^{-x^4/4}.
\]
Integrating from $x$ to infinity and imposing $p_*(x)\phi'(x)\to0$ gives, for $x\ge0$,
\begin{align*}
 e^{-x^4/4}\phi'(x)
 &=\int_x^\infty t e^{-t^4/4}dt\nonumber\\
 &=\frac{\sqrt\pi}{2}\operatorname{erfc}\!\left(\frac{x^2}{2}\right).
\end{align*}
The derivative is therefore
\begin{equation}
 \phi'(x)=\frac{\sqrt\pi}{2}
 e^{x^4/4}\operatorname{erfc}\!\left(\frac{x^2}{2}\right).
 \label{eqS:phi_prime}
\end{equation}
Stationarity and reversibility imply
\[
\mathcal I
 =\langle x\phi\rangle_*
 =-\langle\phi\mathcal L\phi\rangle_*
 =\langle (\phi')^2\rangle_*.
\]
Using evenness of $(\phi')^2$ and Eq.~\eqref{eqS:phi_prime},
\begin{equation}
 \mathcal I
 =\frac{\pi}{2Z}\int_0^\infty
 e^{x^4/4}\operatorname{erfc}^2\!\left(\frac{x^2}{2}\right)dx
 =0.4874954944\ldots .
 \label{eqS:Iintegral}
\end{equation}

\subsection{Critical observation process}

Write $x_N(u)=m(u/\omega_N)/\epsilon_N$ for the finite rescaled
magnetization; its limit is the diffusion $x$ in
Eq.~\eqref{eqS:signal_final}. Over a critical interval of length $O(\omega_N^{-1})$, the integral
of $m$ has size $\epsilon_N/\omega_N=O(N^{1/4}/\kappa)$.
The fast part has variance of order $1/(\kappa\omega_N)$, so its
standard deviation has the same size. Both remain in the rescaled
integrated record
\begin{equation}
 Q_t=\int_0^t z_1(r)dr,\qquad
 y_N(u)=\kappa c_s^{1/4}N^{-1/4}Q_{u/\omega_N}.
 \label{eqS:y_def}
\end{equation}
For finite $N$, retaining $Q_t$ at every time retains the tagged
path, since its slope is $z_1(t)$ between jumps. The limiting channel
describes these complete integrated paths on a fixed interval in $u$.
The leading two-state rates in Eq.~\eqref{eqS:telegraph_rate} have
centered covariance $(1-m^2)e^{-\kappa t}$ at fixed $m$. Hence their
additive-functional variance rate is
\begin{equation}
 \sigma_Q^2(m)=2\int_0^\infty(1-m^2)e^{-\kappa t}dt
 =\frac{2(1-m^2)}\kappa=\frac2\kappa+O(N^{-1/2}).
 \label{eqS:telegraph_variance}
\end{equation}
For the evolving system, apply the flip generator directly to $z_1$:
\[
 Q_hz_1=\kappa(m-z_1)+e_N,\qquad
 |e_N|\le C\kappa(m^2+|h|+N^{-1}).
\]
If $M_z(t)=z_1(t)-z_1(0)-\int_0^t(Q_hz_1)(r)dr$, then
\[
 Q_t=\int_0^tm(r)dr+\frac{M_z(t)}\kappa
      -\frac{z_1(t)-z_1(0)}\kappa
      +\frac1\kappa\int_0^te_N(r)dr.
\]
Multiply by $\kappa c_s^{1/4}N^{-1/4}=\kappa\sqrt{c_s}\epsilon_N$.
The boundary is $O(\epsilon_N)$, and the moment bound gives
\[
 \epsilon_N\mathbb E\int_0^{U/\omega_N}|e_N(r)|dr
 \le\frac{C_U\kappa\epsilon_N}{\omega_N}
       (\epsilon_N^2+|h|+N^{-1})=O_U(\epsilon_N).
\]
Thus, uniformly on a fixed critical interval,
\[
 y_N(u)=\rho_s\int_0^ux_N(r)dr
          +\sqrt{c_s}\epsilon_NM_z(u/\omega_N)+o_{L^1}(1),
 \qquad \rho_s=c_s^{-1/2}.
\]
The observation martingale has jumps bounded by
$2\sqrt{c_s}\epsilon_N$, and its bracket satisfies
\[
 c_s\epsilon_N^2\int_0^{u/\omega_N}4w_1(r)dr
 =2u+o_{L^1}(1).
\]
Let $M_{x,N}$ be the martingale part of $x_N$.
Only tagged flips enter the cross bracket, giving
\[
 \left\langle M_{x,N},\sqrt{c_s}\epsilon_NM_z(\cdot/\omega_N)\right\rangle_u
 =\frac{4\sqrt{c_s}}N\int_0^{u/\omega_N}w_1(r)dr
 =O_U(N^{-1/2}).
\]
The vanishing jumps and these brackets give independent limiting
Brownian noises. Therefore
\begin{equation}
 dy=\rho_sx\,du+\sqrt2\,dV_u,
 \label{eqS:obs_final}
\end{equation}
with the signal in Eq.~\eqref{eqS:signal_final}.
In physical units, the same limit gives the coarse-grained equation
\begin{equation}
 dQ_t\simeq m(t)dt+\sqrt{2/\kappa}\,dB_t.
 \label{eqS:Q_homog}
\end{equation}

The limiting fluctuations contain both the collective signal and
the accumulated tagged noise. To obtain the information in this
channel, the field derivative must be taken at a fixed observed
record. Comparing that likelihood with the microscopic flip
likelihood will then determine the information of the original
tagged trajectory.

\subsection{Information in the limiting channel}

For an observation strength $\rho>0$, the parameter enters the
hidden drift and the stationary density
$p_\vartheta(x)\propto e^{-x^4/4+\vartheta x}$. Their contributions
to the complete signal score are
\begin{align}
 B_U
 &=\left.\partial_\vartheta\log p_\vartheta(x_0)\right|_0
   +\frac12\int_0^U(dx_u+x_u^3du)\nonumber\\
 &=x_0+\frac{W_U}{\sqrt2}
  =\frac{x_0+x_U+\int_0^Ux_u^3du}{2},\qquad
 \mathbb E B_U^2=V_x+U/2.
 \label{eq:complete-signal-score}
\end{align}

The conditional law of the observation given the signal has no
$\vartheta$ dependence. Averaging the complete score over signal
paths compatible with the observation therefore gives
\[
 S_U=\mathbb E[B_U\mid\mathcal Y_U],\qquad
 \mathcal Y_U=\sigma(y_{[0,U]}),\qquad
 F_\vartheta(U)=\mathbb E S_U^2\le V_x+U/2.
\]
The complete signal contributes information at rate $1/2$; the
observation retains the part of its score that can be predicted
from $y$.

For sequential likelihood evaluation, the useful prediction is
the mean of the next observation increment. Write
$\pi_u^{(\vartheta)}(f)=\mathbb E_\vartheta[f(x_u)\mid\mathcal Y_u]$,
and let $\pi_u=\pi_u^{(0)}$. The predicted increment has mean
$\rho\pi_u^{(\vartheta)}(x)du$ and variance $2du$. Its parameter
sensitivity determines the observed score. At a fixed observation
path, define
\begin{equation}
 q_u(f)=\left.\partial_\vartheta
 \pi_u^{(\vartheta)}(f)\right|_{\vartheta=0}.
 \label{eqS:qdef}
\end{equation}

At $\vartheta=0$, subtracting the predicted drift gives the
normalized prediction error
\begin{equation}
 d\mathsf I_u=\frac{dy_u-\rho\pi_u(x)du}{\sqrt2}.
 \label{eqS:innovation}
\end{equation}

The process $\mathsf I_u$ is Brownian under the observation
filtration. The observation log likelihood and its derivative obey
\[
 d\ell_\vartheta
 =\frac{\rho}{2}\pi^{(\vartheta)}(x)\,dy
  -\frac{\rho^2}{4}\pi^{(\vartheta)}(x)^2du.
\]
Differentiating at a fixed $y$ makes the score proportional to the
prediction error:
\begin{align}
 dS_u
 &=\frac{\rho}{2}q_u(x)
 [dy_u-\rho\pi_u(x)du]\nonumber\\
 &=\frac{\rho}{\sqrt2}q_u(x)d\mathsf I_u.
 \label{eqS:score_universal}
\end{align}

Thus the martingale isometry gives the finite-time information and
its rate,
\[
 F_\vartheta(U)=\frac{\rho^2}{2}
                  \int_0^U\mathbb E[q_u(x)^2]\,du,
\]
\begin{equation}
 \mathcal J(\rho)=\lim_{U\to\infty}\frac{F_\vartheta(U)}U
 =\lim_{U\to\infty}\frac{\rho^2}{2U}
                  \int_0^U\mathbb E[q_u(x)^2]\,du.
 \label{eqS:J_filter}
\end{equation}

When the filter and its derivative are stationary, this is
$\rho^2\mathbb E_{\rm inv}[q(x)^2]/2$. The bound on the conditional
score gives $\mathcal J(\rho)\le1/2$.

To compute the prediction and its derivative, let
$\mathcal Lf=-x^3f'+f''$. Conditional averaging of the signal
evolution, followed by the observation update, gives the
Kushner--Stratonovich equation
\begin{equation}
 d\pi(f)=\pi(\mathcal Lf)du
 +\frac{\rho}{\sqrt2}\operatorname{Cov}_\pi(f,x)d\mathsf I.
 \label{eqS:KS}
\end{equation}

The stationary initial state fixes
$\pi_0(f)=\langle f\rangle_*$ and
$q_0(f)=\operatorname{Cov}_*(f,x)$.
The parameter derivative acts both on the signal generator and
on the prediction subtracted in the innovation:
\[
 \left.\partial_\vartheta\mathcal L_\vartheta f\right|_0=f',
 \qquad
 \left.\partial_\vartheta d\mathsf I^{(\vartheta)}\right|_0
       =-\frac{\rho}{\sqrt2}q(x)\,du.
\]
Differentiating the filter equation then gives
\begin{align}
 dq(f)={}&\left[
 q(\mathcal Lf)+\pi(f')
 -\frac{\rho^2}{2}\operatorname{Cov}_\pi(f,x)q(x)
 \right]du\nonumber\\
 &+\frac{\rho}{\sqrt2}
 \left[q(fx)-q(f)\pi(x)-\pi(f)q(x)\right]d\mathsf I.
 \label{eqS:tangent_filter}
\end{align}

These equations propagate the conditional distribution and its
field derivative for a one-dimensional hidden signal. Only the
predicted magnetization sensitivity $q(x)$ enters the information
rate \eqref{eqS:J_filter}.

\subsection{Finite-interval likelihoods}

We now take $\rho=\rho_s$. Throughout the comparison with the spin
process, $F_\vartheta(U)$ and $G(U)$ refer to this channel.
Let $F_N(U)$ be the $\vartheta$ information in the stationary tagged
record of physical duration $U/\omega_N$. Changing the field
coordinate is exact at every $N$:
\begin{equation}
 F_h^{\rm loc}(T)=\alpha_N^2F_N(\omega_NT),\qquad
 J_N=\lim_{U\to\infty}\frac{F_N(U)}U
    =\frac{R_{\rm loc}}{\alpha_N^2\omega_N}.
 \label{eq:physical-record-information}
\end{equation}

Since $\alpha_N^2\omega_N=N\kappa(1+s)^2$, the claimed physical rate
amounts to $J_N\to\mathcal J(\rho_s)$. Squaring the field
amplification contributes $N^{3/2}$, while the slow critical clock
contributes $N^{-1/2}$. The remaining coefficient is the information
per unit critical time.

A fixed critical interval supplies two useful experiments: the
tagged observation alone, and the same observation with the hidden
states at its boundaries supplied. For the latter, define
\begin{equation}
 \begin{aligned}
 G_N(U)&=\mathbb E F_\vartheta
       (z_{1,[0,U/\omega_N]},K_{U/\omega_N}\mid z_1(0),K_0),\\
 G(U)&=\mathbb E_*F_\vartheta(y_{[0,U]},x_U\mid x_0).
 \end{aligned}
 \label{eq:endpoint-information}
\end{equation}

The initial state is held fixed when differentiating the conditional
likelihood. If a long record is divided into blocks and all boundary
states are supplied, the conditional block scores are orthogonal.
For example,
\[
 F_N(mU)\le F_\vartheta(z_1(0),K_0)+mG_N(U).
\]
Thus $G_N(U)/U$ bounds the stationary rate from above. Separated
observation blocks give the complementary lower bound below.
Both comparisons require finite-interval limits:
\begin{equation}
 F_N(U)\longrightarrow F_\vartheta(U),\qquad
 G_N(U)\longrightarrow G(U)
 \quad(U\text{ fixed}).
 \label{eq:fixed-window-information}
\end{equation}

The microscopic record contains the individual flip times.
Expanding its likelihood identifies which of their field-dependent
contributions survive on the critical scale.
With $n=N-1$, the instantaneous tagged flip rate is
$\lambda_z=w_z(h+J\bar m)$, where
\begin{equation*}
 \bar m=\frac1n\sum_{i\ne1}z_i,\qquad
 \bar x=\frac{\bar m}{\epsilon_N},\qquad
 |\bar x-x_N|\le\frac{2}{N\epsilon_N}.
\end{equation*}
The observed spin also changes the hidden flip rates through
$Jz_1/n$. Deleting this term makes the hidden count autonomous, so
that the tagged likelihood can be conditioned on a signal whose
dynamics does not depend on the observed path. The comparison count
has rates
\eqref{eq:count-rates}, $N=n$, and coupling
$J(n-1)/n=J_c+O(N^{-1})$. The coupling shift vanishes on the
critical scale. Since the fields stay bounded and the flip rates are
smooth and positive, the modified hidden rates $\bar w_i$ satisfy
\begin{equation*}
 \eta_i=\frac{w_i-\bar w_i}{\bar w_i}=O(N^{-1}).
\end{equation*}
Over physical time $U/\omega_N$, the compensated-jump likelihood has
variance bounded by
\begin{equation*}
 \sum_{i\ne1}\mathbb E\int_0^{U/\omega_N}\bar w_i\eta_i^2\,dt
 =O_U(N^{3/2}N^{-2})=O_U(N^{-1/2}).
\end{equation*}
Including the stationary initial weights, let $D_N$ be the
likelihood ratio of the original joint law to this autonomous-count
comparison. Under the comparison law, the same compensated-jump
estimate gives
\begin{equation}
 \|D_N-1\|_{L^p}=O_U(N^{-1/4}),\qquad
 \|\partial_\vartheta D_N\|_{L^2}=o_U(1)
 \quad(p,U\text{ fixed}).
 \label{eq:autonomy-error}
\end{equation}
For the derivative, the corresponding rate and compensator bounds are
\begin{equation*}
 \partial_\vartheta\log\frac{w_i}{\bar w_i}
 =O((N\alpha_N)^{-1}),\qquad
 O_U\left(\frac{1}{\omega_N\alpha_N}\right)=O_U(N^{-1/4}).
\end{equation*}
The initial tagged weight differs from a fair sign by
$O(\epsilon_N(1+|\bar x|))$, as does its $\vartheta$ derivative.
The stationary moment bound \eqref{eq:stationary-tails} controls both.

The autonomous-count experiment retains the signal-dependent
tagged intensity $\lambda_z=w_z(h+J\bar m)$. To expand its
conditional likelihood, use a product reference: the same hidden
count, independent of a fair telegraph record with rate $\kappa/2$.
Let $\Lambda_N$ be the conditional tagged likelihood relative to
this fair record. The original joint density relative to the product
reference is then $D_N\Lambda_N$. The hidden law still depends on
$\vartheta$; its score will supply the surviving parameter response.
At zero field, the tagged intensity ratio is
\begin{equation*}
 \frac{\lambda_z}{\kappa/2}
 =1-z\epsilon_N\bar x
  +\frac{s}{(1+s)^2}\epsilon_N^2\bar x^2
  +O(\epsilon_N^3|\bar x|^3+N^{-1}).
\end{equation*}
If $C_t$ counts tagged flips, $dz_1=-2z_1\,dC$ gives
\begin{align}
 y_N^0(u)
 &=-2\sqrt{c_s}\epsilon_N
     \int_0^{u/\omega_N}z_1(t)[dC_t-\kappa\,dt/2]\nonumber\\
 &=y_N(u)+\sqrt{c_s}\epsilon_N
       [z_1(u/\omega_N)-z_1(0)].
 \label{eq:reference-record}
\end{align}
Under the reference law,
\begin{equation*}
 \langle y_N^0\rangle_u=2u,\qquad
 \sup_u|\Delta y_N^0(u)|\le2\sqrt{c_s}\epsilon_N\longrightarrow0.
\end{equation*}
Thus $y_N^0$ tends to variance-two Brownian motion, independently of
the hidden count. The boundary term in Eq.~\eqref{eq:reference-record}
vanishes as well.

Write $\eta_z=\lambda_z/(\kappa/2)-1$. Expanding the tagged log
likelihood to second order gives
\begin{align}
 \log\Lambda_N
 &=\int\log(1+\eta_z)\,dC-\frac\kappa2\int\eta_z\,dt\nonumber\\
 &=\int\eta_z(dC-\kappa\,dt/2)
    -\frac\kappa4\int\eta_z^2\,dt+o_{L^2}(1)\nonumber\\
 &=\frac{\rho_s}{2}\int_0^U\bar x_u\,dy_N^0(u)
   -\frac{\rho_s^2}{4}\int_0^U\bar x_u^2\,du+o_{L^2}(1).
 \label{eq:record-likelihood}
\end{align}
For the local estimates, restrict the hidden path to $[-M,M]$
throughout $[0,U]$. Moment bounds below remove this restriction.
The discarded martingale variance and drift are
\begin{equation*}
 O_{M,U}(\epsilon_N^4/\omega_N)=o(1),\qquad
 O_{M,U}(\epsilon_N^3/\omega_N)=o(1).
\end{equation*}
The same bounds hold after differentiation in $\vartheta$.
The limiting conditional likelihood is consequently
\begin{equation}
 \Lambda(x,y^0)=\exp\left[
  \frac{\rho_s}{2}\int_0^Ux_u\,dy_u^0
  -\frac{\rho_s^2}{4}\int_0^Ux_u^2\,du\right].
 \label{eq:limiting-likelihood}
\end{equation}

Integrating this limiting likelihood over the signal gives the
observed likelihood. The complete score
\eqref{eq:complete-signal-score} differentiates that signal law, so
the likelihood and its score numerator are
\begin{equation}
 Z_{\rm obs}=\mathbb E_x\Lambda,\qquad
 Z_{\rm obs}'=\mathbb E_x(\Lambda B_U),\qquad
 F_\vartheta(U)=\mathbb E^{\rm ref}
                  \frac{(Z_{\rm obs}')^2}{Z_{\rm obs}}.
 \label{eq:observed-information}
\end{equation}

Here $\mathbb E_x$ integrates over the stationary zero-field
signal law, and $\mathbb E^{\rm ref}$ averages over the independent
variance-two Brownian record. The prime is the derivative at
$\vartheta=0$.

To identify the finite-count score, let $k$ count the positive hidden
spins. Denote its upward and downward rates by
$\bar\Lambda_k,\bar M_k$, and its stationary weights by $\bar P_k$.
Its spacing is $\bar d_N=2/(n\epsilon_N)$.
On $|\bar x|\le M$, the jump scores are
\begin{equation*}
 \begin{aligned}
 \partial_\vartheta\log\bar\Lambda_k
 &=\frac{\bar d_N}{2}+O_M(\epsilon_N\bar d_N),\\
 \partial_\vartheta\log\bar M_k
 &=-\frac{\bar d_N}{2}+O_M(\epsilon_N\bar d_N).
 \end{aligned}
\end{equation*}
Summing the compensated jumps and adding the stationary initial score
gives
\begin{equation*}
 B_{N,U}
 =\frac{\bar x_0+\bar x_U+\int_0^U\bar x_u^3du}{2}+o_{L^2}(1).
\end{equation*}
The error variance is $O_M(U\epsilon_N^2)$.
The direct tagged-rate score has variance $O_U(N^{-1})$ and also
vanishes. In the autonomous-count comparison, the observed likelihood
and its derivative are
\[
 Z_N=\mathbb E_{\bar x}\Lambda_N,\qquad
 Z_N'=\mathbb E_{\bar x}
       (\Lambda_N B_{N,U}+\partial_\vartheta\Lambda_N).
\]
The first term differentiates the hidden law; the second
differentiates the tagged likelihood at a fixed hidden path.
The latter has vanishing information by the preceding estimate.
Thus the quantities that must converge are the averages of
$\Lambda_N$ and $\Lambda_N B_{N,U}$ over hidden paths.

Replace the stochastic integral in Eq.~\eqref{eq:record-likelihood}
by its left-endpoint sum on a mesh of size $\eta$.
The increment bound and the martingale isometry give
\begin{equation*}
 \begin{aligned}
 \mathbb E|\bar x_{u+\eta}-\bar x_u|^2
   &\le C_M(\eta+\bar d_N^2),\\
 \mathbb E|\log\Lambda_N-\log\Lambda_N^\eta|^2
   &\le C_{M,U}(\eta+\bar d_N^2).
 \end{aligned}
\end{equation*}
At fixed $\eta$, the likelihood depends only on the hidden mesh
coordinates and the reference observation increments. Restrict the
latter to a compact set. Integrating over the hidden coordinates
now involves finite-dimensional distributions, so diffusion
convergence gives the limiting mesh likelihood and score numerator.
They are continuous in the observation increments, and the averaged
likelihood is strictly positive on this compact set. The information
integrand $(Z_N')^2/Z_N$ therefore converges as well.

On the bounded signal window, likelihood moments turn the logarithmic
mesh estimate into $L^2$ control of $\Lambda_N$ and
$\Lambda_NB_{N,U}$. Averaging over hidden paths contracts these
errors. We can therefore let $\eta\downarrow0$ in $Z_N,Z_N'$ and
their information integral. The moment estimates below then remove
the observation and signal cutoffs.

For $G_N(U)$, the final count is itself observed. A lattice
probability at that count must therefore converge, after division
by the lattice spacing, to the corresponding transition density.
For this calculation, include both boundary states in the joint
likelihood. Its information is the stationary initial-state
information plus $G_N(U)$; the limit is $V_x+G(U)$.
The complete scores $B_{N,U}$ and $B_U$ can then differentiate
these joint likelihoods. The initial contribution will be
subtracted at the end.

On $[-M,M]$, stop both processes at exit and impose zero boundary
values. Let $(\lambda_{N,r},\phi_{N,r})$ and
$(\lambda_r,\phi_r)$ be the eigenpairs of their negative killed
generators in critical time, orthonormal in their respective
stationary weights. The gradient energies satisfy
\begin{equation*}
 \mathcal E_N(f)
 =\frac1{\omega_N}\sum_k\bar P_k\bar\Lambda_k
       (f_{k+1}-f_k)^2
 \longrightarrow\int_{-M}^Mp_*(x)f'(x)^2\,dx.
\end{equation*}
Linear interpolation of a normalized eigenfunction of bounded
energy obeys
\begin{equation*}
 |f(y)-f(x)|^2\le |y-x|\int_{-M}^M|f'(r)|^2\,dr.
\end{equation*}
The interpolated eigenfunctions thus have uniformly convergent
subsequences. Summation by parts in their equations gives the
diffusion eigenvalue equation with zero boundary values. Sampling its
eigenfunctions in the discrete variational formula supplies the
reverse bound. For each fixed $r$,
\begin{equation*}
 \lambda_{N,r}\to\lambda_r,\qquad
 \phi_{N,r}\to\phi_r.
\end{equation*}
Comparison with the interval Laplacian and the same derivative bound
give
\begin{equation*}
 \lambda_{N,r}\ge c_Mr^2,\qquad
 \|\phi_{N,r}\|_\infty^2\le C_M(1+\sqrt{\lambda_{N,r}}).
\end{equation*}
The kernel expansions consequently have uniformly summable tails for
every positive $u$. If $P_N^M$ and $p^M$ denote the killed kernels,
\begin{align}
 \frac{P_N^M(u;\xi,\zeta)}{\bar d_N}
 &=\frac{\bar P_N(\zeta)}{\bar d_N}
   \sum_r e^{-\lambda_{N,r}u}\phi_{N,r}(\xi)\phi_{N,r}(\zeta)\nonumber\\
 &\longrightarrow p_*(\zeta)
   \sum_r e^{-\lambda_ru}\phi_r(\xi)\phi_r(\zeta)
   =p^M(u;\xi,\zeta).
 \label{eq:transition-density}
\end{align}
Products of these kernels give the joint coordinates and final state
on a fixed mesh. Weighting by $\Lambda$ and $\Lambda B_U$ gives the
endpoint likelihood and its derivative; the preceding mesh bound
then removes the discretization.

The stationary tail bound and martingale moment bounds give, at fixed $U$,
\begin{equation*}
 \sup_N\mathbb E B_{N,U}^4\le C_U,\qquad
 \sup_N\mathbb E\sup_{u\le U}|\bar x_u|^4\le C_U.
\end{equation*}
Let $E_M$ be the event that the signal stays in $[-M,M]$.
Restricting the likelihood and its score numerator to $E_M$ changes
the information integral by at most
\begin{equation*}
 C_U(\mathbb E B_U^4)^{1/2}\Pr(E_M^c)^{1/2}\longrightarrow0,
\end{equation*}
uniformly in $N$, with $B_{N,U}$ used for the count experiment.
This follows by splitting on $E_M,E_M^c$ and applying conditional
Cauchy--Schwarz; revealing the indicator bounds the information lost
when the pieces are combined. On bounded signal windows, likelihood
moments remove the observation cutoff in the same way.

Equation~\eqref{eq:autonomy-error} transfers these limits from the
autonomous-count comparison to the original spin process.
Its stationary initial-state information tends to $V_x$.
Subtracting that contribution from the joint endpoint information
gives $G_N(U)\to G(U)$, with the initial state held fixed in the
conditional likelihood. Together with the observation-only
calculation, this proves both limits in
Eq.~\eqref{eq:fixed-window-information}.

\subsection{Long-time information rate}

Use $U$ for a block duration, $L$ for an unobserved gap, and $\tau$
for the total duration, all in critical-time units. Retaining
separated blocks reduces their dependence; supplying the hidden
state between contiguous blocks removes that dependence from the
conditional scores. These two operations bound the information rate
using the finite experiments just considered.

First apply the block construction to the limiting channel.
Let $\gamma>0$ be a lower bound on its relaxation gap; the count
estimate below also gives this bound for the rescaled finite-$N$
chains. Retain channel observations of length $U$, separated by gaps
of length $L$, and let $s_j$ be the zero-parameter score of block $j$.
As a fixed statistic, its variance and parameter response are
\begin{equation*}
 \mathbb E s_j^2=F_\vartheta(U),\qquad
 \left.\partial_\vartheta\mathbb E_\vartheta s_j\right|_0
 =F_\vartheta(U).
\end{equation*}
The gap bounds the covariance of separated blocks:
\begin{equation*}
 |\operatorname{Cov}(s_j,s_{j+r})|
 \le F_\vartheta(U)e^{-\gamma[L+(r-1)(U+L)]}.
\end{equation*}
For $X_m=\sum_{j=1}^m s_j$, summation gives
\begin{equation*}
 \begin{aligned}
 \left.\partial_\vartheta\mathbb E_\vartheta X_m\right|_0
    &=mF_\vartheta(U),\\
 \operatorname{Var}X_m&\le mF_\vartheta(U)b(U,L),\qquad
 b(U,L)=1+\frac{2e^{-\gamma L}}{1-e^{-\gamma(U+L)}}.
 \end{aligned}
\end{equation*}
Applying the score inequality \eqref{eqS:CS_FI} yields
\begin{equation*}
 F_\vartheta(m(U+L))\ge\frac{mF_\vartheta(U)}{b(U,L)}.
\end{equation*}

For an upper bound, use contiguous blocks and reveal all hidden
boundary states. Their conditional scores have zero conditional
mean and are mutually orthogonal. Each contributes $G(U)$, so
\begin{equation*}
 F_\vartheta(mU)
 \le F_\vartheta(y_{[0,mU]},x_0,x_U,\ldots,x_{mU})
 =V_x+mG(U).
\end{equation*}
After division by the duration, the two inequalities give
\begin{equation}
 \frac{F_\vartheta(U)}{(U+L)b(U,L)}
 \le\liminf_{\tau\to\infty}\frac{F_\vartheta(\tau)}\tau
 \le\limsup_{\tau\to\infty}\frac{F_\vartheta(\tau)}\tau
 \le\frac{G(U)}U.
 \label{eq:block-bounds}
\end{equation}
The same bounds hold for $F_N,G_N$, with the same $\gamma$ and
$b(U,L)$; the finite initial-state information again disappears after
division by the duration.

The required relaxation estimate follows from the count weights.
For the count component, choose a median $K_{\rm med}$ and write
\begin{equation*}
 R_j=\sum_{i=K_{\rm med}}^{j-1}\frac1{P_i\Lambda_i}
 \quad(j>K_{\rm med}),
\end{equation*}
with the positive sum over $j\le i<K_{\rm med}$ on the left.
Weighted Cauchy--Schwarz along the count edges gives
\begin{align}
 (f_j-f_{K_{\rm med}})^2
 &\le R_j\sum_iP_i\Lambda_i(f_{i+1}-f_i)^2,\nonumber\\
 \operatorname{Var}_P f
 &\le\left(\sum_jP_jR_j\right)
       \sum_iP_i\Lambda_i(f_{i+1}-f_i)^2.
 \label{eq:count-gap-bound}
\end{align}
In $|m|\le C\epsilon_N$, $P_K\asymp d_N$ and
$\Lambda_K\asymp N\kappa$. Outside this interval, summing the decreasing
ratios gives
\begin{equation*}
 P_jR_j\le\frac{C}{N\kappa m_j^3}
 \quad(C\epsilon_N<m_j<m_0).
\end{equation*}
Beyond a fixed $m_0>0$, the ratios are geometric and
$P_jR_j\le C/(N\kappa)$. The left tail is identical by symmetry.
The central part and the two tails therefore give
\begin{equation*}
 \sum_jP_jR_j
 \le\frac{C}{\omega_N}
    +\frac{C}{\kappa}\int_{\epsilon_N}^{m_0}m^{-3}\,dm
    +\frac{C}{\kappa}
 \le\frac{C'}{\omega_N}.
\end{equation*}
Equation~\eqref{eq:count-gap-bound} gives a count gap at least
$\gamma\omega_N$. The contrast loss \eqref{eq:contrast-generator}
gives a gap at least $c\kappa$ in the other part.
Thus the tagged-count chain has gap at least $\gamma\omega_N$, with
$\gamma>0$ independent of $N$. Passing the energy inequality to the
diffusion gives its gap at least $\gamma$ in critical time.

It remains to compare a long observed block with the same block
and its endpoints. The excess $V_x+G(\tau)-F_\vartheta(\tau)$ is precisely
the information carried by those endpoints conditional on the record.
A terminal state can inform the early score only through the memory
that survives the last $L$ units of observation. The recent score has
variance $L/2$. Splitting the score into these two parts will make
the endpoint cost sublinear in the total duration.

We need the final-state information in two experiments: one
with a stationary random initial state, and one conditioned on the
initial state. Their complete scores are $B_\tau$ and
$M_\tau=W_\tau/\sqrt2$, respectively. For a complete-path statistic
$Z$, the changes in its conditional prediction when $x_\tau$ is
supplied are
\begin{equation*}
 \begin{aligned}
 D_{\tau}Z&=\mathbb E[Z\mid\mathcal Y_\tau,x_\tau]
         -\mathbb E[Z\mid\mathcal Y_\tau],\\
 D_\tau^{(0)}Z&=\mathbb E[Z\mid\mathcal Y_\tau,x_0,x_\tau]
         -\mathbb E[Z\mid\mathcal Y_\tau,x_0].
 \end{aligned}
\end{equation*}
Orthogonal projection gives the final-state increments for the
stationary experiment and the experiment with known initial state:
\begin{equation*}
 C_{\rm stat}(\tau)=\mathbb E|D_{\tau}B_\tau|^2,\qquad
 C_{\rm known}(\tau)=\mathbb E|D_\tau^{(0)}M_\tau|^2.
\end{equation*}

Conditioning on the path up to $\tau-L$ can affect the final score
only through the hidden state at $\tau-L$. The expected difference
between terminal filters started from that state and from stationarity
is measured by
\begin{equation*}
 \delta_L=\mathbb E\left\|
  \Pr(x_L\in\cdot\mid x_0,y_{[0,L]})
  -\Pr(x_L\in\cdot\mid y_{[0,L]})\right\|_{\rm TV},
 \qquad \|\mu-\nu\|_{\rm TV}=\sup_A|\mu(A)-\nu(A)|.
\end{equation*}
The quartic signal has strictly positive transition densities and
converges to stationarity in total variation. Its Gaussian observation
likelihood is positive on finite intervals. Filter stability therefore
gives $\delta_L\to0$.

For an early-path statistic $Z$, determined by the path up to $\tau-L$
and bounded by $K$, the Markov property gives
\begin{equation*}
 \mathbb E|D_{\tau}Z|^2+\mathbb E|D_\tau^{(0)}Z|^2\le CK^2\delta_L.
\end{equation*}
For this bound, use the filter started from $p_*$ at $\tau-L$ and
conditioned on the last $L$ observations as a common reference.
Given the early path, the terminal filter starts at $x_{\tau-L}$ and
has expected distance $\delta_L$ from that reference. Without the
early path it is a conditional mixture of these filters, whose
expected distance is no larger. The triangle inequality gives
$2\delta_L$. Averaging over the stationary initial state gives the
same estimate when $x_0$ is known.
For $\mathbb E Z^4\le C$, truncate at $K$ and use
\begin{equation*}
 \mathbb E[Z^2\mathbf1_{|Z|>K}]\le\frac{\mathbb E Z^4}{K^2},
 \qquad
 \mathbb E|D_{\tau}Z|^2+\mathbb E|D_\tau^{(0)}Z|^2
 \le C(K^2\delta_L+K^{-2}).
\end{equation*}
The early scores are unbounded, but division by $\sqrt \tau$ keeps
their fourth moments bounded independently of the total duration:

\begin{equation*}
 \mathbb E B_\tau^4\le C(1+\tau^2),\qquad
 \mathbb E M_\tau^4\le C\tau^2,\qquad
 \mathbb E|B_\tau-B_{\tau-L}|^2
 =\mathbb E|M_\tau-M_{\tau-L}|^2=L/2.
\end{equation*}
Apply the truncation bound to $B_{\tau-L}/\sqrt \tau$ and
$M_{\tau-L}/\sqrt \tau$, then add the last $L$ increments.
The choice $K=\delta_L^{-1/4}$ balances the bounded-score error
and the truncation tail, giving
\begin{equation}
 \frac{C_{\rm stat}(\tau)+C_{\rm known}(\tau)}\tau
 \le C\left(\sqrt{\delta_L}+\frac{L}{\tau}+\frac1\tau\right).
 \label{eq:endpoint-score-bound}
\end{equation}
For $\delta_L=0$, take $K\to\infty$.
Letting $\tau\to\infty$ and then $L\to\infty$ makes both endpoint
increments $o(\tau)$.

Stationary reversibility transfers the terminal-state estimate
to the initial state. Since it holds for every nearby $\vartheta$,
\begin{equation*}
 F_\vartheta(y,x_0)=F_\vartheta(y,x_\tau).
\end{equation*}
The information chain rule now yields
\begin{align}
 V_x+G(\tau)
 &=F_\vartheta(y,x_0,x_\tau)\nonumber\\
 &=F_\vartheta(\tau)+C_{\rm stat}(\tau)+C_{\rm known}(\tau),\nonumber\\
 G(\tau)-F_\vartheta(\tau)
 &=C_{\rm stat}(\tau)+C_{\rm known}(\tau)-V_x=o(\tau).
 \label{eq:endpoint-negligible}
\end{align}
Choose $U$ along a sequence attaining the lower long-time limit of
$F_\vartheta(U)/U$. The upper bound in Eq.~\eqref{eq:block-bounds}
and Eq.~\eqref{eq:endpoint-negligible} force its upper long-time limit
to be the same. This proves the existence of \eqref{eqS:J_filter}.
At fixed $N$, filter forgetting follows directly from the positive
hidden flip rates: on an interval with no tagged flip, the conditional
hidden propagator is a strictly positive matrix. Its normalized action
contracts differences of posterior distributions. Such intervals have
probability bounded below and recur. Score moments are
$O_N(1+\tau^2)$, so the same endpoint calculation gives
\begin{equation*}
 G_N(\tau)-F_N(\tau)=o_N(\tau),\qquad
 J_N=\lim_{\tau\to\infty}\frac{F_N(\tau)}\tau.
\end{equation*}

At fixed $U,L$, insert \eqref{eq:fixed-window-information} into the
finite-$N$ block bounds \eqref{eq:block-bounds}:
\begin{equation}
 \frac{F_\vartheta(U)}{(U+L)b(U,L)}
 \le\liminf_NJ_N\le\limsup_NJ_N\le\frac{G(U)}U.
 \label{eq:rate-convergence}
\end{equation}
Choose $L=\sqrt U$. Equations~\eqref{eqS:J_filter} and
\eqref{eq:endpoint-negligible} give
\begin{equation*}
 \frac{L}{U}\to0,\qquad b(U,L)\to1,\qquad
 \frac{G(U)}U-\frac{F_\vartheta(U)}U\to0.
\end{equation*}
Both bounds in Eq.~\eqref{eq:rate-convergence} tend to
$\mathcal J(\rho_s)$, proving
\begin{equation*}
 J_N\longrightarrow\mathcal J(\rho_s).
\end{equation*}

The rate at each fixed $N$ was defined with physical observation
time tending to infinity. In the last comparison $U,L$ were held
fixed while $N\to\infty$; only afterward was $U$ increased.
Using $\alpha_N^2\omega_N=N\kappa(1+s)^2$, we obtain the asymptotic local Fisher-information rate:
\begin{align}
 R_{\rm loc}
 &=\alpha_N^2\omega_NJ_N\nonumber\\
 &=N\kappa(1+s)^2\mathcal J(\rho_s)+o(N\nu)\nonumber\\
 &=N\nu(1-s)(1+s)^2\mathcal J(\rho_s)+o(N\nu).
 \label{eqS:Rloc_proved}
\end{align}

\subsection{Information fraction and evaluation}

For the complete record, each spin contributes its own compensated
jump score. If $C_i$ counts the flips of site $i$, differentiating
the joint likelihood gives
\begin{equation*}
 S_h^{\rm full}(T)=\partial_h\log\pi_h(\bm z_0)
  +\sum_i\int_0^T\partial_h\log w_i\,[dC_i-w_i\,dt].
\end{equation*}
The compensated integrals are mutually orthogonal and orthogonal to
the initial score. Their expected quadratic variation gives
\begin{equation*}
 F_h^{\rm full}(T)
 =F_h(\bm z_0)+T\sum_{\bm z}\pi_h(\bm z)
                   \sum_iw_i(\partial_h\log w_i)^2.
\end{equation*}
The initial term is finite for fixed $N$. Dividing by $T$ and taking
$T\to\infty$ therefore yields
\begin{equation}
 R_{\rm full}
 =\sum_{\bm z}\pi_h(\bm z)
   \sum_{i=1}^Nw_i(\bm z;h)
      [\partial_h\log w_i(\bm z;h)]^2,
 \label{eqS:full_FI_general}
\end{equation}
the continuous-record information formula.

At zero cavity field, $w_i=\kappa/2$ and
$\partial_h a_i=1$. The logarithmic rate derivative is
\begin{align}
 \left.\partial_h\ln w_i\right|_{a_i=0}
 &=\left[-\frac{z_i\operatorname{sech}^2a_i}{1-z_i\tanh a_i}
 -\frac{s z_i\operatorname{sech}^2a_i}{1-sz_i\tanh a_i}
 \right]_{a_i=0}\nonumber\\
 &=-z_i(1+s).
 \label{eqS:dlogw}
\end{align}
A single site therefore contributes $\kappa(1+s)^2/2$ to the
information-rate density at zero cavity field.

At criticality the cavity fields satisfy $\mathbb E|a_i|=O(\epsilon_N)$.
On their bounded range, the rate integrand is smooth, so
\[
 \left|w_i(\partial_h\log w_i)^2-\frac\kappa2(1+s)^2\right|
 \le C\kappa|a_i|.
\]
Taking its stationary average and summing the sites gives
\begin{align}
 R_{\rm full}
 &=N\frac\kappa2(1+s)^2+o(N\nu)\nonumber\\
 &=\frac{N\nu}{2}(1-s)(1+s)^2+o(N\nu).
 \label{eqS:Rfull_proved}
\end{align}

Combining Eqs.~\eqref{eqS:Rloc_proved} and \eqref{eqS:Rfull_proved},
\begin{equation}
 \lim_{N\to\infty}\frac{R_{\rm loc}}{R_{\rm full}}
 =2\mathcal J(\rho_s)
 \label{eqS:ratio_proved}
\end{equation}
with $\rho_s=c_s^{-1/2}$.

In the limiting channel,
$\partial_\vartheta\langle x\rangle|_0=V_x$. The mean response and
variance of the integrated observation are therefore
\[
 \left.\partial_\vartheta\frac{\mathbb E y_U}{U}\right|_0=\rho V_x,
 \qquad \operatorname{Var}_0(y_U/U)
 =\frac2U(1+\rho^2\mathcal I)+o(U^{-1}).
\]
The score inequality gives
\begin{equation}
 \mathcal J(\rho)\ge
 \frac{\rho^2V_x^2}{2(1+\rho^2\mathcal I)}.
 \label{eqS:Jlower}
\end{equation}
Conditional expectation of the complete score gives the complementary
bound $F_\vartheta(U)\le V_x+U/2$, hence $\mathcal J(\rho)\le1/2$.
At $\rho=\rho_s$, the resulting interval for the retained fraction is
\begin{equation}
 0<\eta_s^{\rm mean}\le2\mathcal J(\rho_s)\le1,
 \qquad \eta_s^{\rm mean}
 =\frac{V_x^2}{c_s+\mathcal I}
 =\frac4{c_s+\mathcal I}
   \left[\frac{\Gamma(3/4)}{\Gamma(1/4)}\right]^2.
 \label{eqS:eta_mean}
\end{equation}

For $s=1/3$, $c_s=3/8$, and
\begin{equation}
 \eta_{1/3}^{\rm mean}=0.5297959050\ldots .
 \label{eqS:eta_num}
\end{equation}

The evaluation quoted in the Letter at $\rho=\sqrt{8/3}$ gives
$\mathcal J\simeq0.266$, corresponding to a local/full fraction near
$0.532$. The analytic bound already guarantees more than half the
complete-record information from the time average of one spin.
The quoted filtering estimate lies close to that bound.

The time-average bound can also be calculated directly at finite
$N$. Both its response and correlation area reduce to linear
equations for the tagged-count generator $Q_h$. Use coordinates
$(z,k)$, where $z=z_1$ and $k=K-(1+z)/2$ counts the positive hidden
spins. The corresponding tagged and hidden cavity fields are
\begin{align}
 a_1&=h+\frac{J}{N-1}[2k-(N-1)],
 \label{eqS:a_tag_finite}\\
 a_-&=h+\frac{J}{N-1}[z+2k-(N-1)+1],
 \label{eqS:a_minus}\\
 a_+&=h+\frac{J}{N-1}[z+2k-(N-1)-1].
 \label{eqS:a_plus}
\end{align}
The transitions $z\mapsto-z$, $k\mapsto k+1$, and $k\mapsto k-1$
have rates $w_z(a_1)$, $(N-1-k)w_-(a_-)$, and $kw_+(a_+)$.
These are the entries of the exact $2N\times2N$ generator $Q_h$
in the $(z,k)$ coordinates.

The stationary row vector and its field derivative obey
\[
 \begin{aligned}
 \bm\pi_hQ_h&=0,&\bm\pi_h\bm1&=1,\\
 \bm\pi_0'Q_0&=-\bm\pi_0Q_0',&\bm\pi_0'\bm1&=0.
 \end{aligned}
\]
Then $\chi_1=\sum_{z,k}z\,\pi_0'(z,k)$.
For $f(z,k)=z$, the zero-field mean vanishes. Solve
\begin{equation}
 -Q_0g=f,\qquad \sum_{z,k}\pi_0(z,k)g(z,k)=0.
 \label{eqS:finite_poisson}
\end{equation}
The correlation area is
\begin{equation}
 \int_0^\infty C_1(t)dt
 =\sum_{z,k}\pi_0(z,k)f(z,k)g(z,k).
 \label{eqS:finite_corr}
\end{equation}
This correlation integral and $\chi_1$ determine the finite-$N$
value of $R_{\rm mean}$.

For the numerical evaluation of Eq.~\eqref{eqS:J_filter}, $x$ is
discretized on a symmetric grid. A nearest-neighbor generator matches
the first two jump moments of
$\mathcal L_\vartheta f=(-x^3+\vartheta)f'+f''$ to the grid
accuracy. The forward filter and its parameter derivative are
propagated and normalized at each Gaussian observation increment.
After burn-in, the score increments give
\[
 \widehat{\mathcal J}=U^{-1}\sum_n(\Delta S_n)^2.
\]
\section{Parameter directions and local information}

The common field drives the collective mode coherently. A change of
parameter need not have this property: spatial components can cancel,
and a perturbation of hidden coordinates can leave the observed law
unchanged. The retained fraction therefore depends on the driven
direction as well as on the dynamics and the observed variable.

\subsection{Spatially varying fields}

Keep the spin dynamics and observe $z_1$, but replace the common
field by $h_j=\theta g_j$. The profile $g\ne0$ is fixed, and
$\theta$ is estimated at zero. Its spatial mean and squared strength
are $\bar g=N^{-1}\sum_jg_j$ and $\|g\|^2=\sum_jg_j^2$.
For a distributed field, $\|g\|^2=\Theta(N)$ and $g_1=O(1)$,
the retained fraction is controlled by
\[
 \Gamma_N=\frac{\bar g^2}{N^{-1}\sum_jg_j^2}.
\]
If $\Gamma_N\to\Gamma$, that fraction tends to
$2\Gamma\mathcal J(\rho_s)$. Thus the mean field must remain
comparable to its root-mean-square strength. A field of equal
strength on a fraction $p$ of the sites gives $\Gamma=p$, even
when the observed site is outside the driven set. A balanced
positive and negative field has $\Gamma_N=0$.

This dependence follows from the symmetry of the observed
experiment. At $\theta=0$, a permutation of the hidden sites leaves
the tagged record unchanged. The derivative of its likelihood is
linear in $g$, so averaging $g$ over these permutations leaves the
observed score unchanged. Only the projected profile
\[
 (P_1g)_1=g_1,\qquad
 (P_1g)_j=\frac{\sum_{k\ne1}g_k}{N-1}\quad(j\ne1)
\]
can affect that score. The full record, in contrast, resolves every
site. Orthogonality of its compensated jump scores gives
\[
 R_{\rm full}(g)=r_N\|g\|^2,\qquad
 r_N=\mathbb E_0\!\left[
 w_1(\partial_{a_1}\log w_1)^2\right]
 =\frac\kappa2(1+s)^2+o(\nu).
\]

The projected field contains a uniform part and a contrast at the
observed site. With $v=e_1-\bm1/N$ and $\bm1=(1,\ldots,1)$,
\[
 P_1g=\bar g\bm1+\frac{g_1-\bar g}{1-1/N}v.
\]
These components are orthogonal in the full information norm.
Their respective fractions of the original squared field strength
are
\begin{equation}
 \Gamma_N(g)=\frac{N\bar g^2}{\|g\|^2},\qquad
 \ell_N(g)=\frac{(g_1-\bar g)^2}{(1-1/N)\|g\|^2},\qquad
 \eta_N=\frac{R_{\rm loc}(\bm1)}{Nr_N}.
 \label{eq:field-overlaps}
\end{equation}
Relative to the original full-record budget, the uniform part gives
\[
 \frac{R_{\rm loc}(\bar g\bm1)}{R_{\rm full}(g)}
 =\Gamma_N\eta_N.
\]
The contrast contributes at most its full information, since
$R_{\rm loc}(v)\le r_N\|v\|^2$. The reverse triangle inequality
for the two observed scores gives the lower bound below. For the
upper bound, use
\[
 R_{\rm loc}(g)=R_{\rm loc}(P_1g)
 \le R_{\rm full}(P_1g)
 =r_N\|g\|^2(\Gamma_N+\ell_N).
\]
Together,
\begin{equation}
 \left(\sqrt{\Gamma_N\eta_N}-\sqrt{\ell_N}\right)_+^2
 \le\frac{R_{\rm loc}(g)}{R_{\rm full}(g)}
 \le\Gamma_N+\ell_N,\qquad x_+=\max(x,0).
 \label{eq:field-fraction-bounds}
\end{equation}
Expanding the squared norm of the sum of the two local scores
also gives
\[
 \left|\frac{R_{\rm loc}(g)}{R_{\rm full}(g)}
                   -\Gamma_N\eta_N\right|
 \le2\sqrt{\Gamma_N\eta_N\ell_N}+\ell_N.
\]
Consequently,
\begin{equation}
 \ell_N\to0:\qquad
 \liminf_N\frac{R_{\rm loc}(g)}{R_{\rm full}(g)}>0
 \quad\Longleftrightarrow\quad
 \liminf_N\Gamma_N(g)>0.
 \label{eq:spatial-field-criterion}
\end{equation}
For the distributed profiles specified above,
$\ell_N=O(N^{-1})$ and $\eta_N\to2\mathcal J(\rho_s)$.
This proves the mean-to-root-mean-square criterion and the limit
$2\Gamma\mathcal J(\rho_s)$.

A vanishing fraction need not mean exactly zero finite-size
information: the tagged-site contrast can still contribute.
If $g_1=0$ and the hidden fields sum to zero, however, $P_1g=0$.
The observed likelihood then has zero first derivative at every
$N$ and every record length, so its Fisher information is exactly
zero there.

\subsection{Gaussian networks}

For a linear network, the driven and observed directions can be
compared without permutation symmetry. Consider
\begin{equation}
 d\bm Z=-\kappa A\bm Z\,dt+\kappa\bm b\theta\,dt
                  +\sqrt{2\kappa D}\,d\bm W_t,\qquad
 Y=\bm c^{\mathsf T}\bm Z.
 \label{eq:gaussian-network}
\end{equation}
Here the scalar coordinate $Y$ is recorded directly; $\bm c=e_i$
selects one site. The matrix $D$ is symmetric positive definite,
all eigenvalues of $A$ have positive real part, and
$\bm b,\bm c\ne0$. The calculation allows nonsymmetric $A$.

The stationary mean is $A^{-1}\bm b\theta$. Subtracting it gives
$\bm\xi=\bm Z-A^{-1}\bm b\theta$ with dynamics
\[
 d\bm\xi=-\kappa A\bm\xi\,dt+\sqrt{2\kappa D}\,d\bm W_t.
\]
The centered Gaussian path law is independent of the parameter.
Thus a drive affects the entire observed law only through the
mean $\bm c^{\mathsf T}A^{-1}\bm b\theta$. The resulting rates are
\begin{equation}
 R_Y=\frac\kappa2
       \frac{(\bm c^{\mathsf T}A^{-1}\bm b)^2}
            {\bm c^{\mathsf T}A^{-1}DA^{-\mathsf T}\bm c},
 \qquad
 R_{\rm full}=\frac\kappa2\bm b^{\mathsf T}D^{-1}\bm b.
 \label{eq:gaussian-exact-rates}
\end{equation}
In noise-weighted coordinates,
$\bm u=D^{-1/2}\bm b$ is the drive direction and
$\bm w=D^{1/2}A^{-\mathsf T}\bm c$ is the static response direction
selected by the readout. Equation~\eqref{eq:gaussian-exact-rates}
gives
\begin{equation}
 \frac{R_Y}{R_{\rm full}}
 =\frac{(\bm u^{\mathsf T}\bm w)^2}{|\bm u|^2|\bm w|^2}.
 \label{eq:gaussian-direction-criterion}
\end{equation}
For a size-dependent network, a nonzero fraction survives precisely
when this squared overlap stays bounded away from zero.

The exact local rate follows from matching lower and upper
bounds. Set $\bm v=A^{-\mathsf T}\bm c$. Integrating
$\bm v^{\mathsf T}d\bm Z$ gives
\[
 \int_0^T Y_t\,dt
 =\bm v^{\mathsf T}\bm b\,\theta T
  -\frac{\bm v^{\mathsf T}(\bm Z_T-\bm Z_0)}\kappa
  +\sqrt{\frac2\kappa}\bm v^{\mathsf T}D^{1/2}\bm W_T.
\]
The stationary boundary term has bounded variance, and its
covariance with the martingale is $o(T)$ by Cauchy--Schwarz.
The variance per unit time therefore tends to
$2\bm v^{\mathsf T}D\bm v/\kappa$. The score inequality for the
time average gives
\[
 R_Y\ge\frac\kappa2
       \frac{(\bm v^{\mathsf T}\bm b)^2}
            {\bm v^{\mathsf T}D\bm v}.
\]
The complete dynamic score is
$\sqrt{\kappa/2}\int_0^T\bm b^{\mathsf T}D^{-1/2}d\bm W_t$.
Its variance gives the full rate in
Eq.~\eqref{eq:gaussian-exact-rates}.

For the upper bound, take a drive parallel to $D\bm v$ with
the same observed mean:
\[
 \bm b_\parallel
 =D\bm v\,\frac{\bm v^{\mathsf T}\bm b}
                  {\bm v^{\mathsf T}D\bm v}.
\]
It satisfies $\bm v^{\mathsf T}\bm b_\parallel
=\bm v^{\mathsf T}\bm b$. Its centered law is unchanged, so it
produces exactly the same observed path law. The local information
is bounded by the complete information for this drive,
\[
 R_Y\le\frac\kappa2
       \bm b_\parallel^{\mathsf T}D^{-1}\bm b_\parallel
 =\frac\kappa2
       \frac{(\bm v^{\mathsf T}\bm b)^2}
            {\bm v^{\mathsf T}D\bm v}.
\]
This equals the time-average lower bound and proves
Eq.~\eqref{eq:gaussian-exact-rates}. In particular,
$\bm v^{\mathsf T}\bm b=0$ makes the entire observed law
independent of $\theta$.

For symmetric $A$ and $D=\mathsf I$, the mode expansion is
\[
 A^{-1}e_i=\sum_\alpha
        \frac{u_\alpha(i)}{\lambda_\alpha}\bm u_\alpha.
\]
A mode contributes $u_\alpha(i)^2/\lambda_\alpha^2$ to
$|\bm w|^2$. A small relaxation eigenvalue increases its weight
only at sites where that mode is present. The retained fraction
also requires its overlap with the driven direction. The slow
relaxation rate alone therefore does not determine the information
available at a chosen site.

\subsection{Nonlinear observations}

In the Gaussian family above, the parameter changes only the
observed mean. Nonlinear dynamics can transmit a perturbation
through the variance or temporal correlations while leaving that
mean unchanged. A nonzero response in any finite observation
interval is enough to give a positive long-time information rate
when the covariances of separated blocks are summable, as they
are under the positive-gap condition below.

Consider the stationary channel
\begin{equation}
 d\bm X=b_a(\bm X)du+\sqrt2\,d\bm W,\qquad
 d\bm Y=H(\bm X)du+\sqrt2\,d\bm V,
 \label{eq:general-channel}
\end{equation}
with independent noises and a fixed continuous observation
function $H$. Let $\pi_a$ be the stationary density,
$\sigma_\pi=\partial_a\log\pi_a|_0$, and
$g=\partial_ab_a|_0$. The complete score and rate are
\begin{equation}
 B_U=\sigma_\pi(\bm X_0)+\frac1{\sqrt2}\int_0^U
                  g(\bm X_u)\cdot d\bm W_u,\qquad
 \mathbb E B_U^2=\mathbb E_\pi\sigma_\pi^2+U\mathcal B,\qquad
 \mathcal B=\frac12\mathbb E_\pi|g|^2.
 \label{eq:general-budget}
\end{equation}
Assume that the nearby diffusion family is reversible, that its
zero-parameter dynamics has a positive gap $\gamma$ and converges
to $\pi$ in total variation, and that
$0<\mathcal B<\infty$ and
$\mathbb E B_U^4\le C(1+U)^2$.

Let $A$ be a bounded statistic of a record of length $U$, with
response $d_A=\partial_a\mathbb E_aA|_0$ and variance $v_A>0$.
Repeat this statistic in blocks separated by time $L$. For $m$
blocks, the response of their sum is $m d_A$, while the gap gives
variance at most $m v_A b(U,L)$, with $b(U,L)$ as in
Eq.~\eqref{eq:block-bounds}. The score inequality yields
\[
 \mathcal J_a(H)\ge
 \frac{d_A^2}{v_A(U+L)b(U,L)},\qquad
 \mathcal J_a(H)=\lim_{\tau\to\infty}\frac{F_a(\tau)}{\tau}.
\]
Thus a mean, variance, or correlation response of a finite record
can establish positive local information without computing its
optimal estimator.

The rate in this bound exists under the stated conditions.
Indeed, the observed score is
$S_U=\mathbb E[B_U\mid\bm Y_{[0,U]}]$.
Writing
$G_a(U)=\mathbb E_\pi F_a(\bm Y_{[0,U]},\bm X_U\mid\bm X_0)$,
the block comparison gives
\[
 \frac{F_a(U)}{(U+L)b(U,L)}
 \le\liminf_{\tau\to\infty}\frac{F_a(\tau)}{\tau}
 \le\limsup_{\tau\to\infty}\frac{F_a(\tau)}{\tau}
 \le\frac{G_a(U)}U.
\]
The last $L$ units of complete score have variance $L\mathcal B$.
Total-variation ergodicity and the positive Gaussian observation
likelihood give filter forgetting. The fourth-moment estimate
therefore controls the early score exactly as in
Eq.~\eqref{eq:endpoint-score-bound}. Reversibility transfers the
terminal-state estimate to the initial state, giving
$G_a(U)-F_a(U)=o(U)$. Choosing $L=\sqrt U$ makes the upper and
lower rate limits agree.

A nonzero derivative of a finite observed path law has density
$S_U\ne0$ and hence pairs nontrivially with some bounded statistic
$A$. Conversely, if every such derivative vanishes, every
$F_a(U)$ is zero. The finite-record response test is therefore also
necessary:
\begin{equation}
 \frac{\mathcal J_a(H)}{\mathcal B}>0
 \quad\Longleftrightarrow\quad
 \exists\,U>0:\quad
 \left.\partial_a\Pr_a(\bm Y_{[0,U]}\in\cdot)\right|_0\ne0.
 \label{eq:general-criterion}
\end{equation}

For a scalar signal observed through $H(x)=\rho x$, $\rho\ne0$,
every nonzero drift perturbation is visible in this sense,
including those with zero mean response. More generally, this
holds whenever $H$ is injective. To see why Gaussian observation
noise cannot erase a signed change of the signal law, suppose the
observed derivative vanishes on an interval of length $U$.
For every deterministic vector-valued step function $f$,
\[
 0=\mathbb E\!\left[B_Ue^{i\int_0^Uf\cdot d\bm Y}\right]
  =e^{-\int_0^U|f|^2du}
    \mathbb E\!\left[B_Ue^{i\int_0^Uf\cdot H(\bm X_u)du}\right].
\]
The Gaussian factor is nonzero. Uniqueness of Fourier transforms
then makes the signed derivative of the law of
$\int_0^\cdot H(\bm X_u)du$ vanish. Continuity of $H(\bm X_u)$
and injectivity of $H$ identify that law with the hidden path law:
the time integral determines $H(\bm X_u)$, which determines
$\bm X_u$. Its derivative would also vanish, contradicting
$\mathbb E B_U^2>0$ when $g\ne0$ in $L^2(\pi)$.
This argument concerns the distinguishability of path laws;
it does not require recovering a hidden realization from a noisy
measurement.

For a readout that is not injective, the finite-record response
test distinguishes visible and invisible perturbations. Consider
two coupled coordinates with only $x_1$ observed:
\begin{equation}
 \begin{aligned}
 V_{a,b}(x_1,x_2)&=\frac{x_1^4+x_2^4}{4}
       +\frac\beta2x_1^2x_2^2-a x_2+\frac b2x_2^2,
       \qquad\beta>0,\\
 d\bm X&=-\nabla V_{a,b}(\bm X)du+\sqrt2\,d\bm W,\qquad
 dY=\rho x_1du+\sqrt2\,dV,\quad\rho>0.
 \end{aligned}
 \label{eq:two-mode-channel}
\end{equation}
Both parameters act on the hidden coordinate $x_2$. Reflection
$x_2\mapsto-x_2$ changes the force $a$ to $-a$ without changing
the observed process. Its law is even in $a$, so at $a=b=0$,
\[
 F_a(Y_{[0,U]})=0\quad\text{for every }U,\qquad
 \mathcal J_a=0,\qquad\mathcal B_a=1/2.
\]
The stiffness $b$ instead changes the even fluctuations of the
hidden coordinate. Through the term $\beta x_1^2x_2^2/2$, it
changes the observed variance.

The stationary density is proportional to $e^{-V_{a,b}}$, so the
stiffness score is $-(x_2^2-\mathbb E x_2^2)/2$ and
\[
 \left.\partial_b\mathbb E_bx_1^2\right|_0
 =-\frac12\operatorname{Cov}(x_1^2,x_2^2).
\]
Conditional on $q=x_2^2$, the density of $x_1$ is proportional to
$e^{-x_1^4/4-\beta qx_1^2/2}$. Consequently
\[
 f(q)=\mathbb E[x_1^2\mid x_2^2=q],\qquad
 f'(q)=-\frac\beta2
       \operatorname{Var}(x_1^2\mid x_2^2=q)<0.
\]
For independent copies $Q,Q'$ of $x_2^2$,
\[
 2\operatorname{Cov}(f(Q),Q)
 =\mathbb E[(f(Q)-f(Q'))(Q-Q')]<0.
\]
Since this is also twice
$\operatorname{Cov}(x_1^2,x_2^2)$, the response
$d=\partial_b\mathbb E_bx_1^2|_0$ is positive. On a short
observation interval,
\[
 \mathbb E_bY_U^2=2U+\rho^2\mathbb E_bx_1^2U^2+O(U^3),\qquad
 \operatorname{Var}_0(Y_U^2)=8U^2+O(U^3).
\]
The variance statistic $Y_U^2$ therefore gives
\begin{equation}
 F_b(U)\ge\frac{\rho^4d^2}{8}U^2[1+O(U)]>0
 \quad(U>0\text{ sufficiently small}).
 \label{eq:two-mode-visible-stiffness}
\end{equation}
This response occurs even though $\mathbb E_bY_U=0$ for every $b$.

For this quartic system, the conditions that turn the finite
response into a positive rate follow from confinement. The
diffusion is reversible with stationary density $e^{-V}/Z$.
Its inward drift and elliptic noise give total-variation
ergodicity. For the stationary gap, set
$\mathcal W=e^{V/2}$ and
$\mathcal L=\Delta-\nabla V\cdot\nabla$. Then
\[
 \Psi=-\frac{\mathcal L\mathcal W}{\mathcal W}
 =|\nabla V|^2/4-\Delta V/2\ge c|\bm x|^6-C,
\]
and integration by parts gives
\[
 0\le\int|\nabla f-f\nabla\log\mathcal W|^2d\pi
 =\int|\nabla f|^2d\pi-\int f^2\Psi\,d\pi.
\]
Choose a ball $B$ outside which $\Psi\ge1$, and subtract the
$\pi$-mean of $f$ inside $B$. The Poincare inequality on the ball
and the lower bound on $\Psi$ inside it give
\[
 \int_Bf^2d\pi\le C_B\int_B|\nabla f|^2d\pi,\qquad
 \int_{B^c}f^2d\pi
 \le\int|\nabla f|^2d\pi+C\int_Bf^2d\pi.
\]
The two bounds control the total variance by the gradient energy,
so the gap is positive. Quartic tails give all polynomial
stationary moments. The force drift derivatives are constant and
the stiffness derivative is $-x_2e_2$; the martingale fourth-moment
bound gives $\mathbb E B_U^4\le C(1+U)^2$.
Fix a sufficiently small $U_0>0$ for which
Eq.~\eqref{eq:two-mode-visible-stiffness} gives $F_b(U_0)>0$.
The block bound then yields
\[
 \mathcal J_b\ge\frac{F_b(U_0)}{(U_0+L)b(U_0,L)}>0,
 \qquad\mathcal B_b=\mathbb E x_2^2/2.
\]
Thus the short-interval variance response produces a positive
long-time rate. The hidden force and hidden stiffness have different
retained fractions in the same dynamics and with the same readout.

For a microscopic family, the physical size of the information
also depends on its critical scales. Suppose
$u=\omega_Nt$, $a=\alpha_N\theta$, the scaled gaps have a uniform
positive lower bound, and the observed and endpoint informations
converge on each fixed interval. If the complete rate divided by
$\alpha_N^2\omega_N$ tends to $\mathcal B$, the same block
comparison gives
\begin{equation}
 \frac{R_{{\rm loc},\theta}}{\alpha_N^2\omega_N}
 \longrightarrow\mathcal J_a(H),\qquad
 \frac{R_{{\rm full},\theta}}{\alpha_N^2\omega_N}
 \longrightarrow\mathcal B
 \quad\Longrightarrow\quad
 \frac{R_{{\rm loc},\theta}}{R_{{\rm full},\theta}}
 \longrightarrow\frac{\mathcal J_a(H)}{\mathcal B}.
 \label{eq:general-microscopic-rate}
\end{equation}
The response test determines whether the fraction is positive;
$\alpha_N^2\omega_N$ determines the scale of the physical rates.
For the mean-field signal, $g=1$ and $\mathcal B=1/2$, recovering
$2\mathcal J(\rho_s)$.

Finally, a force on the observed coordinate gives an explicit
family with an extensive physical information rate. Replace
$-a x_2$ by $-c x_1$. Its response is
\[
 \left.\partial_c\mathbb E_cY_U\right|_0
 =\rho U\mathbb E_0x_1^2\ne0,\qquad\mathcal B_c=1/2,
\]
so $\mathcal J_c>0$. With $V_0=V_{0,0}$, the family
\[
 d\bm m=\kappa[-\nabla V_0(\bm m)+\theta e_1]dt
                +\sqrt{2\kappa/N}\,d\bm W_t,\qquad
 dQ=\rho m_1dt+\sqrt{2/\kappa}\,dV_t
\]
becomes this channel under
$\bm X=N^{1/4}\bm m$, $u=\kappa N^{-1/2}t$,
$c=N^{3/4}\theta$, and $Y=\kappa N^{-1/4}Q$.
The squared field amplification times the critical clock is
$(N^{3/4})^2\kappa N^{-1/2}=N\kappa$. Hence
\begin{equation}
 R_{{\rm loc},\theta}=N\kappa\mathcal J_c,\qquad
 R_{{\rm full},\theta}=N\kappa/2,\qquad
 R_{{\rm loc},\theta}/R_{{\rm full},\theta}=2\mathcal J_c>0.
 \label{eq:coupled-extensive-rate}
\end{equation}
The two-mode calculation separates the origin of the extensive
scale from the visibility of the perturbation: the critical
rescaling fixes the scale, while the observed response determines
whether its coefficient is nonzero.

\section{Sparse critical readout}
\label{secS:sparse}

This section proves the sparse-readout result of the Letter. We use
Secs.~S1--S3 without repeating their critical-scaling, complete-record,
filter-stability, or block-comparison proofs. The new ingredients are
the fixed-$k$ joint observation experiment, a relaxation check for its
additional tagged coordinates, and matching bounds on the information
lost through a noisy observation of the quartic signal.

\subsection{Statement and order of limits}

Keep $h=0$, $J=J_c$, $0\le s<1$, and $\nu>0$ fixed, with the
rapid-readout dynamics of Eq.~\eqref{eqS:rate}. For a predetermined
set $A$ of $k$ distinct sites, define
\begin{equation}
 R_{N,k}=\lim_{T\to\infty}\frac1T
 F_h\bigl((z_i(t))_{i\in A,\,0\le t\le T}\bigr),
 \qquad 1\le k\le N.
 \label{eqS:sparse_rate_def}
\end{equation}
The records are stationary and include all flip times. Permutation
symmetry makes the rate independent of the choice of $A$. As in S2,
the physical-time limit is taken first at each fixed $N$.

\noindent\textbf{Theorem (sparse critical readout).}
For every fixed positive integer $k$,
\begin{align}
 R_{N,k}
 &=N\kappa(1+s)^2\mathcal J\!\left(\sqrt{k/c_s}\right)
   +o_k(N\nu),
 \label{eqS:sparse_micro_rate}\\
 \eta_{k,s}\equiv\lim_{N\to\infty}\frac{R_{N,k}}{R_{N,N}}
 &=2\mathcal J\!\left(\sqrt{k/c_s}\right).
 \label{eqS:sparse_eta}
\end{align}
With $V_x$ as in Eq.~\eqref{eqS:Vx}, the information loss obeys

\begin{equation}
 \frac{9c_sV_x^2}{k+9c_s}
 \ \le\ 1-\eta_{k,s}
 \ \le\ \frac{9c_s}{k-6c_s},\qquad k>6c_s.
 \label{eqS:sparse_two_sided}
\end{equation}

The left inequality holds for every $k\ge1$. In particular,
\begin{equation}
 1-\eta_{k,s}=\Theta(k^{-1}),\qquad
 k_{\min}(\varepsilon;s)\equiv
 \min\{k\in\mathbb N:\eta_{k,s}\ge1-\varepsilon\}
 =\Theta(\varepsilon^{-1}).
 \label{eqS:sparse_complexity}
\end{equation}
Finally, for every sequence $1\le k_N\le N$ with $k_N\to\infty$,
\begin{equation}
 \frac{R_{N,k_N}}{R_{N,N}}\longrightarrow1.
 \label{eqS:sparse_sequence}
\end{equation}
This includes every such sequence with $k_N/N\to0$.
The constants in the fixed-$k$ microscopic comparison may depend on
$k$ and $s$; Eq.~\eqref{eqS:sparse_two_sided} bounds the limiting
fractions, not a uniform finite-size remainder. The theorem does not
assert a sharp coefficient of $k^{-1}$.

\subsection{The fixed-\texorpdfstring{$k$}{k} microscopic observation experiment}

By exchangeability take $A=\{1,\ldots,k\}$ and put $n=N-k$.
For the large-$N$ argument $k$ remains fixed and $n\ge2$.
Write $\boldsymbol\sigma=(z_1,\ldots,z_k)$, let $H$ count the
positive hidden spins, and set
\[
 \bar m=(2H-n)/n,\qquad \bar x=\bar m/\epsilon_N.
\]
The pair $(\boldsymbol\sigma,H)$ is a finite-state Markov chain.
The critical scales $\epsilon_N,\omega_N,\alpha_N$ are exactly
those of Eq.~\eqref{eqS:critical_scaling}, and
$|\bar x-x_N|\le2k/(N\epsilon_N)$.

\emph{Autonomous hidden comparison.}
Delete the contributions of the observed spins from the hidden
cavity fields, and delete the other observed spins from each observed
cavity field. The comparison hidden count is autonomous, with
$n$-spin coupling
\begin{equation}
 J_{N,k}^{\rm hid}=J_c\frac{n-1}{N-1}
 =J_c\left(1-\frac{k}{N-1}\right),
 \label{eqS:sparse_hidden_coupling}
\end{equation}
and the observed spins, conditional on this count path, have independent
flip dynamics with rates
\[
 \bar\lambda_j=w_{\sigma_j}
       \left(h+J_c\frac{n}{N-1}\bar m\right).
\]
Initialize the hidden count in its comparison stationary law and the
$k$ observed signs as independent fair signs. This initialization is
used only for the finite-window comparison; the original experiment
remains stationary.

The coupling displacement in
Eq.~\eqref{eqS:sparse_hidden_coupling} produces a linear drift of
order $kN^{-1/2}\bar x$ in critical time. Its stationary quadratic
potential correction is also $O_k(N^{-1/2}\bar x^2)$.
Consequently the count-weight expansion, quartic tail bounds, and
killed-kernel argument of S1 and S2, in particular
Eqs.~\eqref{eq:stationary-control}, \eqref{eq:stationary-tails},
and \eqref{eq:transition-density}, have the same limits for this
comparison count. It converges to the signal
Eq.~\eqref{eqS:signal_final}, with the same complete signal score.

The deleted field is at most $J_ck/(N-1)$ at any hidden site and
$J_c(k-1)/(N-1)$ at an observed site. Smoothness and strict positivity
of the rates on the bounded cavity-field range therefore give a
relative intensity error $O(k/N)$. On a fixed critical interval
$[0,U]$, the quadratic variations controlling the joint likelihood
comparison are bounded by
\begin{equation}
 O\!\left(n\kappa\frac U{\omega_N}\frac{k^2}{N^2}\right)
   =O_U(k^2N^{-1/2}),\qquad
 O\!\left(k\kappa\frac U{\omega_N}\frac{k^2}{N^2}\right)
   =O_U(k^3N^{-3/2})
 \label{eqS:sparse_autonomy_brackets}
\end{equation}
for hidden and observed flips, respectively.
The stationary-weight expansion gives a comparison-initialization
error $O_{k,p}(\epsilon_N)$ in $L^p$, including its
$\vartheta$ derivative. Applying the compensated-jump estimate
of Eq.~\eqref{eq:autonomy-error} to these bounds, the joint
Radon--Nikodym derivative $D_{N,k}$ satisfies
\begin{equation}
 \|D_{N,k}-1\|_{L^p}=O_{k,U,p}(N^{-1/4}),\qquad
 \|\partial_\vartheta D_{N,k}\|_{L^2}=o_{k,U}(1).
 \label{eqS:sparse_autonomy}
\end{equation}
Here and below $p$ and $U$ are fixed, and the bounds hold on compact
$\vartheta$ intervals. For the derivative estimate, the logarithmic
intensity-ratio derivative is $O(k/(N\alpha_N))$, and its
summed compensator contribution is $O_U(kN^{-1/4})$.
Thus the comparison controls parameter scores as well as path laws.

\emph{Joint conditional likelihood.}
Use as product reference the autonomous hidden count and $k$
independent fair telegraph records. If $C_j$ counts flips of sign
$\sigma_j$, define the martingale reference records
\begin{equation}
 y^0_{N,j}(u)=-2\sqrt{c_s}\epsilon_N
 \int_0^{u/\omega_N}\sigma_j(t)
       [dC_j(t)-\kappa\,dt/2],\qquad j=1,\ldots,k.
 \label{eqS:sparse_reference_records}
\end{equation}
Each differs from the rescaled integrated spin record by the bounded
endpoint term in Eq.~\eqref{eq:reference-record}. Their quadratic
variations are $2u$, their mutual cross variations vanish, and their
jumps tend to zero. Under the product reference they therefore
converge jointly to independent variance-two Brownian records,
independent also of the hidden driving noise.

On a bounded hidden-path window, the intensity expansion used in
Eq.~\eqref{eq:record-likelihood} now holds for each $j$:
\[
 \frac{\bar\lambda_j}{\kappa/2}
 =1-\sigma_j\epsilon_N\bar x
  +\frac{s}{(1+s)^2}\epsilon_N^2\bar x^2
  +O_k\bigl(\epsilon_N^3(1+|\bar x|^3)+N^{-1}\bigr).
\]
Conditional independence lets us add, rather than multiply out,
the $k$ log likelihoods. The result is
\begin{equation}
 \log\Lambda_{N,k}
 =\frac{\rho_s}{2}\sum_{j=1}^k\int_0^U\bar x_u\,dy^0_{N,j}(u)
   -\frac{k\rho_s^2}{4}\int_0^U\bar x_u^2\,du+r_{N,k},
 \qquad
 \|r_{N,k}\|_{L^2}+\|\partial_\vartheta r_{N,k}\|_{L^2}
 \longrightarrow0.
 \label{eqS:sparse_likelihood}
\end{equation}
The discarded martingale variances and drift bounds are finite sums
of the corresponding errors below Eq.~\eqref{eq:record-likelihood},
namely $O_{k,M,U}(\epsilon_N^4/\omega_N)$ and
$O_{k,M,U}(\epsilon_N^3/\omega_N)$ on $|\bar x|\le M$.
The direct field score in the $k$ observed intensities has variance
$O_{k,U}(\kappa/(\omega_N\alpha_N^2))=O_{k,U}(N^{-1})$,
so it vanishes. The surviving score comes from the hidden count
and converges to $B_U$ in Eq.~\eqref{eq:complete-signal-score}.

It follows that the limiting joint channel is
\begin{equation}
 dx=(-x^3+\vartheta)du+\sqrt2\,dW_u,\qquad
 dy_j=\rho_sx\,du+\sqrt2\,dV_{j,u},\quad j=1,\ldots,k,
 \label{eqS:sparse_vector_channel}
\end{equation}
with independent $W,V_1,\ldots,V_k$.
This assertion also holds at the level of Fisher information:
let $F_{N,k}(U)$ be the $\vartheta$ information in the observed
records of duration $U/\omega_N$, and define $G_{N,k}(U)$ by
also supplying the hidden endpoint $H_{U/\omega_N}$ and holding
$(\boldsymbol\sigma(0),H_0)$ fixed when differentiating, exactly
as in Eq.~\eqref{eq:endpoint-information}. Then
\begin{equation}
 F_{N,k}(U)\longrightarrow F_k(U),\qquad
 G_{N,k}(U)\longrightarrow G_k(U)
 \quad (U\text{ and }k\text{ fixed}),
 \label{eqS:sparse_fixed_window}
\end{equation}
where $F_k,G_k$ denote the corresponding channel experiments.
To apply the existing proof of
Eq.~\eqref{eq:fixed-window-information}, replace its single
observation mesh by a $k$-component mesh and use
Eq.~\eqref{eqS:sparse_likelihood}. The mesh-error estimates are
finite sums, the complete-score fourth moments remain bounded,
and the quartic tails remove the hidden-path cutoff. The hidden
coordinate is still a single count, so the endpoint-kernel
convergence is precisely the count argument of
Eq.~\eqref{eq:transition-density}, with the vanishing coupling
shift noted above. Finally Eq.~\eqref{eqS:sparse_autonomy}
transfers both information limits to the original stationary
experiment. No interchange of infinite-time and diffusion limits
has been used here.

\emph{Reduction to one informative channel.}
An orthogonal transformation of the $k$ observations has first
coordinate $\widetilde y=k^{-1/2}\sum_jy_j$. Thus
\begin{equation}
 d\widetilde y=\sqrt{k}\,\rho_sx\,du+\sqrt2\,d\widetilde V_u.
 \label{eqS:sparse_effective_channel}
\end{equation}
The remaining $k-1$ coordinates are independent Brownian motions,
independent of $(x,\widetilde y)$, and their laws have no parameter
dependence. The information of the entire vector, including the
endpoint experiment, is therefore exactly that of the scalar
channel at strength $\rho=\sqrt{k/c_s}$.

\subsection{Relaxation and transfer to the stationary microscopic rate}

The block comparison in S2 requires a gap uniform in $N$ on the
critical clock. The count gap is already proved there. The following
check controls the extra label coordinates without repeating that
count calculation.

Let $\pi_N$ be the zero-field stationary law of the full spin chain,
let $\mu_K$ be its uniform conditional law at count $K$, and write
\[
 \mathcal E_N(f,f)=\frac12\sum_{\boldsymbol z}\pi_N(\boldsymbol z)
 \sum_{i=1}^N w_i(\boldsymbol z;0)
      [f(\boldsymbol z^{\,i})-f(\boldsymbol z)]^2
\]
for its Dirichlet form. Put $\bar f(K)=\mathbb E_{\mu_K}f$.
Uniform edge weights between consecutive count layers and Jensen's
inequality imply that the count Dirichlet form of $\bar f$ is
bounded by $\mathcal E_N(f,f)$. The count-gap bound of S2 therefore
gives
\begin{equation}
 \operatorname{Var}_{P}\bar f
 \le\frac{C}{\omega_N}\mathcal E_N(f,f).
 \label{eqS:sparse_count_component_gap}
\end{equation}

For the within-layer variance, the uniform slice satisfies
\begin{equation}
 \operatorname{Var}_{\mu_K}f
 \le\frac1{2N}\mathbb E_{\mu_K}
       \sum_{i<j}[f(\boldsymbol z^{\,ij})-f(\boldsymbol z)]^2,
 \label{eqS:sparse_slice_gap}
\end{equation}
where $\boldsymbol z^{\,ij}$ exchanges coordinates $i$ and $j$.
For completeness, its constant follows from the transposition
generator $\mathcal T_Kf=\sum_{i<j}(f(\boldsymbol z^{\,ij})-f)$.
Writing $q_i=(1+z_i)/2$ and $f_B=\prod_{i\in B}q_i$ with
$|B|=\ell$, direct summation gives
\[
 \mathcal T_K f_B=(K-\ell+1)\sum_{i\in B}f_{B\setminus\{i\}}
                 -\ell(N-\ell+1)f_B.
\]
The polynomial spaces of increasing degree are invariant. They span
all slice functions by degree $K$, or by degree $N-K$ after replacing
$q_i$ by $1-q_i$. Hence the nonzero eigenvalues of $-\mathcal T_K$
are $\ell(N-\ell+1)$, with
$1\le\ell\le\min(K,N-K)$, and its gap is $N$.
This proves Eq.~\eqref{eqS:sparse_slice_gap}; the extreme slices
$K=0,N$ have zero variance.

A nontrivial exchange can be implemented by two spin flips,
$\boldsymbol z\to\boldsymbol z^{\,i}\to
\boldsymbol z^{\,ij}$. The rates are bounded below by
$c\kappa$ and above by $C\kappa$, uniformly in $N$, because the
cavity fields are bounded. Detailed balance also bounds the ratio
of stationary weights across either flip by a constant. Using
$(a+b)^2\le2a^2+2b^2$ along each two-flip path and summing
Eq.~\eqref{eqS:sparse_slice_gap} over $K$ gives
\begin{equation}
 \mathbb E_P\operatorname{Var}_{\mu_K}f
 \le\frac C\kappa\mathcal E_N(f,f).
 \label{eqS:sparse_label_gap}
\end{equation}
Indeed, a given flip edge is used by at most $N$ choices of the
partner site, cancelling the factor $1/N$ in
Eq.~\eqref{eqS:sparse_slice_gap}.
Combining the conditional-variance decomposition with
Eqs.~\eqref{eqS:sparse_count_component_gap} and
\eqref{eqS:sparse_label_gap} gives
\begin{equation}
 \operatorname{Var}_{\pi_N}f
 \le C\left(\omega_N^{-1}+\kappa^{-1}\right)\mathcal E_N(f,f)
 \le\frac{C'}{\omega_N}\mathcal E_N(f,f).
 \label{eqS:sparse_full_gap}
\end{equation}
Restriction to functions of $(\boldsymbol\sigma,H)$ proves the
required gap for every fixed-$k$ observation experiment.

All remaining hypotheses of the S2 block comparison have now been
checked: Eq.~\eqref{eqS:sparse_fixed_window} supplies the two
finite-window information limits, and
Eq.~\eqref{eqS:sparse_full_gap} supplies the uniform gap.
The limiting channel reduces to Eq.~\eqref{eqS:sparse_effective_channel},
so the endpoint-negligibility result
Eq.~\eqref{eq:endpoint-negligible} applies at its fixed observation
strength. At fixed $N$, the strictly positive hidden-count
propagator on an interval with no observed flips gives the same
filter forgetting as in S2; reversibility is inherited from the
full spin process. Applying Eqs.~\eqref{eq:rate-convergence}
and \eqref{eq:general-microscopic-rate}, rather than reproving
their block argument, yields
\[
 \lim_{N\to\infty}\frac{R_{N,k}}{\alpha_N^2\omega_N}
 =\mathcal J\!\left(\sqrt{k/c_s}\right).
\]
The exact identity $\alpha_N^2\omega_N=N\kappa(1+s)^2$ and
the already proved complete-record rate
Eq.~\eqref{eqS:Rfull_proved} establish
Eqs.~\eqref{eqS:sparse_micro_rate} and \eqref{eqS:sparse_eta}.

\subsection{A lower bound on the missing information}

It remains to bound the loss in the scalar channel with an arbitrary
fixed strength $\rho>0$. Let $F_\rho(U)$ be its stationary
$\vartheta$ information on $[0,U]$, and
$\mathcal Y_U=\sigma(y_{[0,U]})$. The complete score and its
conditional projection are already given in S2. In its notation,
\begin{equation}
 D_\rho(U)\equiv V_x+U/2-F_\rho(U)
 =\mathbb E\operatorname{Var}(B_U\mid\mathcal Y_U),
 \qquad
 B_U=x_0+W_U/\sqrt2.
 \label{eqS:sparse_missing_score}
\end{equation}
We use this identity without repeating the score derivation.
Integration by parts against $p_*(x)$ gives
$\mathbb E_*x^4=1$ and $\mathbb E_*x^6=3V_x$.

Introduce an auxiliary shift $a$, not an additional unknown in the
sensing problem, through
\[
 dX^{(a)}=-(X^{(a)}-a)^3du+\sqrt2\,dW_u,\qquad
 dy=\rho(X^{(a)}-a)du+\sqrt2\,dV_u,
 \qquad X^{(a)}_0\sim p_*(\,\cdot-a).
\]
Since $X^{(a)}-a$ has the original stationary signal law, the
observed law is independent of $a$. Its joint score at $a=0$ is
\begin{equation}
 G_U=x_0^3+\frac3{\sqrt2}\int_0^U x_u^2\,dW_u
                 -\frac\rho{\sqrt2}V_U,
 \qquad \mathbb E[G_U\mid\mathcal Y_U]=0.
 \label{eqS:sparse_shift_score}
\end{equation}
The initial term differentiates $p_*(x_0-a)$; the other two terms
are the drift and observation scores. The polynomial stationary
moments justify these differentiations by localization.
Martingale orthogonality and stationarity give
\begin{equation}
 \mathbb E(B_UG_U)=1+\frac32UV_x,\qquad
 \mathbb E G_U^2=3V_x+\frac U2(9+\rho^2).
 \label{eqS:sparse_shift_moments}
\end{equation}
Because the conditional mean of $G_U$ vanishes, its covariance with
$B_U$ equals its covariance with
$B_U-\mathbb E[B_U\mid\mathcal Y_U]$. Cauchy--Schwarz therefore
bounds the missing information from below:
\begin{equation}
 D_\rho(U)\ge
 \frac{(1+\frac32UV_x)^2}{3V_x+\frac U2(9+\rho^2)}.
 \label{eqS:sparse_loss_lower_finite}
\end{equation}
Using the already established existence of $\mathcal J(\rho)$,
division by $U$ and then $U\to\infty$ give
\begin{equation}
 1-2\mathcal J(\rho)\ge\frac{9V_x^2}{\rho^2+9}.
 \label{eqS:sparse_loss_lower}
\end{equation}
In particular, every finite observation strength loses a strictly
positive fraction of the full information rate.

\subsection{An upper bound on the missing information}

The other direction uses convexity of the conditional path law.
The variance form of the Brascamp--Lieb inequality
states that a density proportional to $e^{-\mathcal A}$ with positive
Hessian obeys
\begin{equation}
 \operatorname{Var}(f)\le
 \mathbb E\bigl[\nabla f^{T}(\nabla^2\mathcal A)^{-1}\nabla f\bigr].
 \label{eqS:sparse_BL}
\end{equation}
We apply this finite-dimensional result to an explicit path
approximation and then remove the approximation.

\emph{Conditional path potential.}
The quartic stationary signal has, relative to variance-two Brownian
path measure with Lebesgue measure for the initial coordinate, the
unnormalized density
\begin{equation}
 \exp\left[-\frac{x_0^4+x_U^4}{8}
           -\int_0^U\Psi(x_u)\,du\right],\qquad
 \Psi(x)=\frac{x^6}{4}-\frac32x^2.
 \label{eqS:sparse_path_prior}
\end{equation}
This follows by applying It\^o's formula to the Girsanov integral
$-\frac12\int x^3dx$ and including the stationary initial density.
Multiplication by the conditional observation likelihood adds
$\rho^2x^2/4$ to the bulk potential and a linear functional of
the path. Thus the relevant bulk potential is
\begin{equation}
 \mathcal W_\rho(x)=\frac{x^6}{4}+\frac{\rho^2-6}{4}x^2,
 \qquad
 \mathcal W_\rho''(x)=\frac{15}{2}x^4+\frac{\rho^2-6}{2}
 \ge a_\rho\equiv\frac{\rho^2-6}{2}.
 \label{eqS:sparse_convexity}
\end{equation}
Assume $\rho^2>6$ for the rest of this subsection.

\emph{Finite-dimensional variance bound.}
Set $u_j=j\delta$, $j=0,\ldots,M$, with $M\delta=U$.
Approximate the integral in Eq.~\eqref{eqS:sparse_path_prior} by
its left-endpoint Riemann sum, retaining the exact Brownian
increments and both endpoint factors. Given this approximate
signal, take the observation drift to be $\rho x_j$ on the
$j$th interval. For observed increments
$\Delta y_j=y_{u_{j+1}}-y_{u_j}$, the posterior negative log density
of $\boldsymbol x=(x_0,\ldots,x_M)$, up to its normalizer, is
\begin{equation}
 \mathcal A_{\delta,y}(\boldsymbol x)
 =\frac{x_0^4+x_M^4}{8}
  +\sum_{j=0}^{M-1}\frac{(x_{j+1}-x_j)^2}{4\delta}
  +\delta\sum_{j=0}^{M-1}\mathcal W_\rho(x_j)
  -\frac\rho2\sum_{j=0}^{M-1}x_j\Delta y_j.
 \label{eqS:sparse_discrete_action}
\end{equation}
The last term is linear and cannot change the Hessian. The other
terms imply
\[
 \boldsymbol v^T\nabla^2\mathcal A_{\delta,y}\boldsymbol v
 \ge a_\rho\delta\sum_{j=0}^{M-1}v_j^2
       +\frac1{2\delta}\sum_{j=0}^{M-1}(v_{j+1}-v_j)^2.
\]
In particular the full Hessian is positive definite, although the
bulk sum omits the terminal coordinate.
For
\[
 A_\delta=\frac\delta2\sum_{j=0}^{M-1}x_j^3,
 \qquad
 \nabla A_\delta=\frac{3\delta}{2}(x_0^2,\ldots,x_{M-1}^2,0),
\]
the inverse-Hessian variational formula and the preceding lower
bound give
\[
 \nabla A_\delta^T(\nabla^2\mathcal A_{\delta,y})^{-1}
       \nabla A_\delta
 \le\frac{9\delta}{4a_\rho}\sum_{j=0}^{M-1}x_j^4.
\]
Here the zero terminal component is important: one can drop the
nonnegative difference term in the variational formula without
introducing an unconstrained linear term at $j=M$.
Applying Eq.~\eqref{eqS:sparse_BL} conditionally, then averaging
over the approximate observation law, gives
\begin{equation}
 \mathbb E_\delta\operatorname{Var}_\delta(A_\delta\mid y)
 \le\frac{9}{2(\rho^2-6)}
          \delta\sum_{j=0}^{M-1}\mathbb E_\delta x_j^4.
 \label{eqS:sparse_discrete_variance}
\end{equation}
Brownian bridges within each observed interval are independent of
the approximate hidden signal, so conditioning on the whole
approximate record gives the same variance as conditioning on its
increments.

\emph{Removal of the time mesh.}
We spell out this step to distinguish the path-law argument from
a formal action calculation. Since $\Psi(x)\ge-\sqrt2$, the
unnormalized prior densities of the approximation are bounded by
\[
 e^{\sqrt2 U}\exp[-(x_0^4+x_U^4)/8]
 \le e^{\sqrt2 U}e^{-x_0^4/8}
\]
relative to the same Brownian path measure. This is integrable,
also after multiplication by any fixed polynomial of the path
supremum. Continuity of Brownian paths and dominated convergence
therefore give weighted $L^1$ convergence of the prior densities,
convergence of their normalizers, and uniform polynomial moments.
Under these priors the piecewise-constant signal $x_j$ converges
to $x_u$ in $L^2(du\,d\mathbb P)$.

For a given hidden path, the relative entropy between the Gaussian
observation laws with drifts $\rho x_j$ and $\rho x_u$ is
\[
 \frac{\rho^2}{4}\int_0^U|x_{\lfloor u/\delta\rfloor\delta}-x_u|^2du.
\]
Its expectation tends to zero. Together with the prior convergence,
this yields total-variation convergence of the joint hidden and
observed laws. Conditional mean-square error is the minimum of
$\mathbb E(A-g(y))^2$ over measurable $g$; for bounded $A$ it is
continuous under this joint convergence by restricting $g$ to the
same bounded range. Truncation and the uniform polynomial moments
extend the conclusion to $A_\delta\to
A_U=\frac12\int_0^U x_u^3du$ in $L^2$.
Consequently Eq.~\eqref{eqS:sparse_discrete_variance} passes to
the exact stationary channel, where $\mathbb E_*x^4=1$:
\begin{equation}
 \mathbb E\operatorname{Var}(A_U\mid\mathcal Y_U)
 \le\frac{9U}{2(\rho^2-6)}.
 \label{eqS:sparse_bulk_variance}
\end{equation}
All limits in this mesh argument are at fixed $U$ and $\rho$.

Finally the existing complete-score formula
Eq.~\eqref{eq:complete-signal-score} gives
$B_U=A_U+(x_0+x_U)/2$.
The endpoint term has second moment at most $V_x$.
The triangle inequality for the $L^2$ residual after conditional
projection therefore yields the finite-time bound
\begin{equation}
 D_\rho(U)\le
 \left[\sqrt{\frac{9U}{2(\rho^2-6)}}+\sqrt{V_x}\right]^2.
 \label{eqS:sparse_loss_upper_finite}
\end{equation}
Dividing by $U$ and taking $U\to\infty$ gives
\begin{equation}
 1-2\mathcal J(\rho)\le\frac9{\rho^2-6},\qquad \rho^2>6.
 \label{eqS:sparse_loss_upper}
\end{equation}
Substitution of $\rho^2=k/c_s$ into
Eqs.~\eqref{eqS:sparse_loss_lower} and
\eqref{eqS:sparse_loss_upper} proves
Eq.~\eqref{eqS:sparse_two_sided} and hence the first part of
Eq.~\eqref{eqS:sparse_complexity}.

\subsection{Readout cost and asymptotically complete sparse sets}

Information is nondecreasing when another node's record is retained.
This is true at every $N$ and every duration, and hence for the
stationary rates and their fixed-$k$ limits. From the lower bound
in Eq.~\eqref{eqS:sparse_two_sided}, any $k$ achieving
$\eta_{k,s}\ge1-\varepsilon$ must satisfy
$k\ge9c_s(V_x^2/\varepsilon-1)$. Conversely, the upper bound
shows that $k\ge6c_s+9c_s/\varepsilon$ suffices. Thus
\begin{equation}
 \max\!\left\{1,
   \left\lceil9c_s\left(\frac{V_x^2}{\varepsilon}-1\right)\right\rceil
      \right\}
 \le k_{\min}(\varepsilon;s)
 \le\left\lceil6c_s+\frac{9c_s}{\varepsilon}\right\rceil,
 \qquad 0<\varepsilon<1.
 \label{eqS:sparse_cost_explicit}
\end{equation}
This establishes the necessary and sufficient
$\Theta(\varepsilon^{-1})$ cost in
Eq.~\eqref{eqS:sparse_complexity}. Its independence of $N$ refers
to the limiting fraction $\eta_{k,s}$; it does not bound the finite
network size needed to approximate that fraction.

To prove Eq.~\eqref{eqS:sparse_sequence}, fix an arbitrary positive
integer $K$. Since $k_N\to\infty$, eventually $k_N\ge K$.
Data processing and exchangeability imply
\[
 1\ge\frac{R_{N,k_N}}{R_{N,N}}
   \ge\frac{R_{N,K}}{R_{N,N}}.
\]
Taking the lower limit in $N$ at fixed $K$ gives
\[
 \liminf_{N\to\infty}\frac{R_{N,k_N}}{R_{N,N}}\ge\eta_{K,s}.
\]
Now let $K\to\infty$ and use
Eq.~\eqref{eqS:sparse_two_sided}. Both bounds become one.
This proves completeness even when $k_N$ grows arbitrarily slowly;
in particular, it allows $k_N/N\to0$ and does not require the
readout sets to be nested across different $N$.
The argument never substitutes $k=k_N$ into the nonuniform
remainder in Eq.~\eqref{eqS:sparse_micro_rate}.

The theorem concerns the Fisher-information rate for the common
weak field in the complete temporal records. It neither reconstructs
individual hidden trajectories nor gives completeness for arbitrary
spatial parameter directions. It also leaves the network-wide
measurement-feedback controller unchanged and makes no claim about
finite-time preparation or a time-average-only readout.
